%% file: isw_v9.tex
\documentclass[useAMS,usenatbib]{mn2e}

\usepackage{graphicx,epsfig}
\usepackage{subfigure}
\usepackage{floatflt}

\input{format}

\def\be{\begin{equation}}
\def\ee{\end{equation}}
\title[A new test for the ISW effect]{Cross-correlating \textit{WMAP}5 with 1.5 million LRGs: a new test for
the ISW effect}
\author[U. Sawangwit et al.]{U. Sawangwit$^1$\thanks{E-mail: utane.sawangwit@durham.ac.uk}, T. Shanks$^1$\thanks{E-mail: tom.shanks@durham.ac.uk},
R.D. Cannon$^2$, S.M. Croom$^3$, Nicholas P. Ross$^{1,4}$ 
\newauthor and  D.A. Wake$^{1,5}$\\ 
$^1$Physics Department, University of Durham, South Road, Durham, DH1 3LE, UK\\
$^2$Anglo-Australian Observatory, PO Box 296, Epping, NSW 1710, Australia\\
$^3$School of Physics, University of Sydney, NSW 2006, Australia\\
$^4$Department of Astronomy and Astrophysics, The Pennsylvania State University, University Park, PA 16802, USA\\
$^5$Department of Astronomy, Yale University, CT 06520, USA}

\begin{document}
\date{Accepted 2009 November 16. Received 2009 November 06; in original form 2009 May 25}

\pagerange{\pageref{firstpage}--\pageref{lastpage}} \pubyear{2009}

\maketitle
\label{firstpage}
\begin{abstract}
We present the cross-correlation of the density map of Luminous Red
Galaxies (LRGs) and the temperature fluctuation in the Cosmic Microwave
Background (CMB) as measured by the five-year Wilkinson Microwave
Anisotropy Probe (\textit{WMAP}) observations. The LRG samples were
extracted from imaging data of the Sloan Digital Sky Survey (SDSS) Data
Release 5 (DR5) based on two previous spectroscopic redshift surveys,
the SDSS--LRG and the  2dF--SDSS LRG and QSO (2SLAQ) surveys designed
to have average redshifts of $z\approx0.35$ and  $z\approx0.55$. In
addition we have added a higher--redshift photometric LRG sample based on 
the selection of the AAOmega LRG redshift survey at $z\approx0.7$. The total LRG sample
thus comprises 1.5 million galaxies, sampling a redshift range of $0.2<z<0.9$
over $\approx7600$ square degrees of the sky, probing a total cosmic
volume of $\approx 5.5 h^{-3}~ \rm{Gpc}^{3}$.

First, we find that the new LRG sample at $z\approx0.7$ shows very little
positive evidence for the Integrated Sachs-Wolfe (ISW) effect. Indeed, the
cross-correlation is negative out to $\approx 1\deg$. The standard $\Lambda$CDM
model is rejected at $\approx$2-3\% significance by the new LRG data.  We then
analyse the previous samples at $z\approx0.35$ and $z\approx0.55$. As found by
other authors, these results appear consistent with the standard ISW model,
although the statistical significance remains marginal. We also reproduce the
same result for the magnitude limited SDSS galaxy samples of \cite{Cabre06}
Taking the $z\approx0.35$ and $z\approx0.55$ LRG results in combination with the
new $z\approx0.7$ sample, the overall result is now more consistent with a null
detection than with the standard $\Lambda$CDM model prediction.

We then performed a new test on the robustness of the LRG ISW detections at
$z\approx0.35$ and $z\approx0.55$. We made 8 rotations through $360\deg$ of the CMB maps with
respect to the LRG samples around the galactic pole. We find that in
both cases there are stronger effects at  angles other than zero. This implies
that the $z\approx0.35$ and $z\approx0.55$  ISW detections may still be subject
to systematic errors which combined with the known sizeable statistical errors
may leave the $z\approx0.35$ and $z\approx0.55$ ISW detections looking
unreliable. We have further made the rotation test on several other samples
where ISW detections have been claimed and find that they also show peaks when
rotated. We conclude that in the samples we have tested the ISW effect may 
be absent and we argue that this result may not be in contradiction with previous results.
 
\end{abstract}
\begin{keywords}
cosmology: observations -- cosmic microwave background -- large-scale structure of Universe.
\end{keywords}

\section{Introduction}

Many observations now suggest that we live in a spatially flat, dark energy
dominated Universe
\citep[e.g.][]{Perlmutter99,Cole05,Tegmark06,Riess07,Dunkley09}. In such a
cosmology, positive correlation between the CMB and large-scale-structure (LSS)
is expected due to the decaying gravitational potential (\citealt{Sachs67}). The
deviation of the CMB temperature in the vicinity of LSS is caused by the
non-vanishing difference in the energy gained and lost by the CMB photons as
they traverse a region of over-- or under--density. By integrating across all
the potential wells along the line of sight from the surface of last scattering,
the primodial fluctuations in the CMB are modified by this effect. This
secondary anisotropy of the CMB is called the Integrated Sachs--Wolfe (ISW)
effect and sometimes known as the late-time ISW effect since the dominance of
dark energy in the cosmic energy budget at the present epoch is believed to be
responsible for the current accelerating expansion and hence the decaying
gravitational potential. For a spatially flat Universe, a detection of the
Integrated Sachs-Wolfe (ISW) effect would provide direct dynamical evidence of
the accelerating expansion unlike the geometrical measurement inferred from
standard candles such as the SNIa.

The SNIa results coupled with the CMB evidence that the Universe is nearly flat,
suggests there exists an exotic form of energy with negative pressure. The exact nature
of this so--called dark energy is not yet known but it already entails many
serious problems. Foremost amongst them are the fine-tuning problem and
the cosmic coincidence problem \citep[e.g.][]{Carroll01,Peebles03}.  

The ISW signal in the CMB--galaxy cross--correlation  is very small, generally
less than one $\mu$K, and very difficult to detect. Previous ISW detections
generally have less than 3$\sigma$ statistical significance. These include the
studies of \cite{Fos03}, \cite{Padmanabhan05b} and \cite{Cabre06} who used  SDSS
galaxies in both photo-z and magnitude limited samples and   the WMAP3 dataset.
Other authors have used X-ray sources \citep{BC04} and NVSS radio sources
\citep{Nolta04}. Of these, it seems that up to now the most significant
detection of the ISW effect comes from the NVSS radio sources at 2.3$\sigma$.
Other authors (eg \citealt {Giannantonio08} and \citealt{Ho08}) have made
compilations of the other results and claimed up to 4$\sigma$ ISW detections in
terms of the overall significance. The only other claims of ISW detections at
high significance are the methods that reduced the galaxy samples to focus only
on regions of high or low underdensity. In particular, 
\cite{Granett08} cross-correlated the positions of $\approx$ 100 superclusters
and voids in the MegaZ--LRG \citep{Collister07} sample and \cite{McEwen07}
employed a similar wavelets method  using radio sources from NVSS.

Here we shall search for the ISW effect by using samples of Luminous Red
Galaxies (LRGs) from the SDSS DR5 dataset. LRGs are the most luminous stellar systems 
in the Universe, residing in the most massive dark matter haloes. Having formed
most of their stars much earlier and over short period of time, the objects
appear red with reasonably uniform spectral energy distributions therefore these
galaxy samples can be selected homogeneously and observed out to greater
distance (or lookback time). Moreover, being massive means that the LRGs are
also a highly biased tracer of the LSS (e.g. \citealt {Ross07},
\citealt{Wake08}). The selection techniques for $z < 0.6$ LRG samples have been
well established in the literature. Many LSS studies have been carried out using
these LRG samples including the claimed detections of the ISW effect (e.g.
\citealt{Cabre06}). The recent spectroscopic survey by \citet{Ross08} has shown
that it is possible to extend the selection technique and hence the LRG sample
out to $z\approx1$. Applying this tested algorithm to the entire SDSS imaging
significantly increases the effective volume and makes these LRGs ideal
probes of large-scale structure.

Our main goal is to detect the ISW signal in the CMB by cross-correlating WMAP5 map 
with the new $\bar{z} \approx 0.7$ LRG sample and to test the detection of the ISW effect
caused by the LRGs at lower redshift ($\bar{z} \approx 0.35,0.55$) as claimed by a
number of authors (e.g. \citealt{Padmanabhan05b}, \citealt{Cabre06}). These studies used
the LRG candidates extracted from the SDSS DR3 or DR4 whilst we
are using DR5, $\approx 50$ per\,cent and 20 per\,cent increase in the area coverage, 
respectively. The larger sky coverage should provide a statistical advantage
over the previous studies. Our new higher redshift LRG sample should also provide
a chance to constrain the evolution if such an effect is indeed detected. Moreover,
a recent study by \cite{Douspis08} suggests that the ISW signal-to-noise can be
optimised if the large-scale tracer probes out to a median redshift of 0.8 but 
there is no further improvement after a redshift of unity.
The claim appears to be supported by the cross-correlation analysis of
\cite{Giannantonio08} in which the signal-to-noise of the ISW detection from the
2 Micron All Sky Survey (2MASS; \citealt{Jarrett00}) is $\approx$ 4--6 times smaller
than from the NRAO VLA Sky Survey (NVSS; \citealt{Condon98}) where $\bar{z} \approx
0.1$ and $0.8$ respectively, despite the fact that the two surveys have similar
sky coverage and sky density $(N_{\rm{NVSS}}\approx 2N_{2\rm{MASS}})$. If this is
true then our higher redshift LRG should be more sensitive to the ISW signal and
will provide even higher significance of detection than previous studies using the
LRGs which currently reach $\approx 2\sigma$ significance at best. 
The new sample therefore presents a fresh opportunity to test one of the most crucial 
manifestations of the accelerating expansion, obtaining independent
confirmation of the geometrical inference of the SNIa result if detected and a
challenge to the current standard picture of the Universe otherwise.

The layout of this paper is as follows. We present the data in
\S \ref{sec:data}. We then outline the theoretical prediction and cross-correlation 
technique employed in this study in \S \ref{sec:theory} and \S \ref{sec:technique}, respectively. 
The results and a range of analyses performed to ensure their robustness are given in 
\S \ref{sec:result} and \S \ref{sec:robustness}. The additional sky rotation tests performed on our 
dataset and selections of previously claimed ISW detections are reported in \S \ref{sec:rotate}.
We then present the discussion and conclusion of our studies in \S \ref{sec:discuss} and \S \ref{sec:conclusion}. 
Throughout this study (unless otherwise stated), we assume a standard $\Lambda$CDM 
cosmology with $\Omega_{\Lambda}=0.7,~\Omega_{\rm{m}}=0.3,
~f_{\rm{baryon}}=0.167,~\sigma_{8}=0.8$ and $H_{0}=100h ~\rm{km~s^{-1}~Mpc^{-1}}$ ($h=0.7$
where necessary).  

\section{Data}
\label{sec:data}

\subsection{CMB Temperature Map-\textit{WMAP}5}
\label{sec:wmap5}

The CMB temperature maps used here are taken from the \textit{WMAP} five-year
data release \citep{Hinshaw09}. The data products are publicly
available\footnote{http://lambda.gsfc.nasa.gov/} in Hierarchical Equal Area
isoLatitude Pixelisation (HEALPix, \citealt{HEALPixref}) format. Although the
\textit{WMAP} observes in five frequency bands, we choose to use only the three
highest-frequency bands, namely, \W at 94 GHz, \V at 61 GHz and \Q at
41 GHz as the CMB anisotropy in these ranges are less susceptible to a
contamination from the foreground anisotropy (i.e. synchrotron and free-free
emission) than the lower frequency counterparts. This enable us to test for any
wavelength dependence in the CMB-galaxy cross-correlation where one expects the
ISW signal to be achromatic. However, we shall concentrate our analysis mainly
on the \W band due to its relatively high resolution compared to the other
bands, $12.'6$ FWHM compared to $19.'8$ for \V and $29.'4$ for \Q band. Despite the
fact that the \V band has lower noise than the \W band (hence often the band of 
choice for this type of analysis), we do not observe any major difference in either the
cross-correlation results or their statistical errors (see Fig. \ref{fig:band}).
We also use the Internal Linear Combination (ILC, \citealt{Gold09}) to further check our results, 
although it should be noted that, according to the \textit{WMAP} team, there could 
be a significant structure in the bias correction map at scales smaller than $\approx 10\deg$ 
\citep{Limon08}.

We shall use the temperature maps at a resolution of $N_{\rm{side}}$=512 (res=9)
which for the whole sky, contains 3\,145\,728 pixels each with an area of 
$\approx$\,49\,$\rm{arcmin}^2$. The foreground-contaminated regions of the sky, mainly in
Galactic Plane and Magellanic Cloud including extragalactic point sources, are
excluded using a combination of `Extended temperature analysis mask' (KQ75, 
\citealt{Gold09}) and `Point source catalogue mask' \citep{Wright09}. After
applying the masks, we are left with 2\,239\,993 pixels ($\approx 70$ per\,cent). 
The maps contain thermodynamic temperatures with the dipole contribution subtracted 
from the data by the \textit{WMAP} team \citep{Hinshaw09}.

\subsection{Luminous Red Galaxies}
\label{sec:lrg}
The Luminous Red Galaxy (LRG) photometric samples are extracted from the SDSS
DR5 \citep{Adelman07} imaging data based on three LRG spectroscopic redshift
surveys whose median redshifts are $\approx 0.35, 0.55~\rm{and} ~0.7$
\citep{Eisenstein01,Cannon06,Ross08}. In essence, these surveys utilised a crude
but effective determination of photometric redshift (photo-z), owing to the strong 4000 \AA\ 
break of a typical E/S0 galaxy spectral energy distribution (SED). 
As the break is redshifted through the SDSS $u,g,r,i,~\rm{and}~z$ bandpasses, 
its colour-colour track exhibits a distinctive turning point at various redshifts 
for different colour pairs. Moreover, their uniform SEDs ensure that they have 
an extremely tight locus in the colour space. This allows the potential LRGs in 
the desired redshift ranges to be selected uniformly using their locations on the 
colour-colour plane coupled with the luminosity threshold set by the appropriate 
magnitude limit. 

\begin{figure}
\hspace{-0.5cm}
\centering
\includegraphics[scale=0.53]{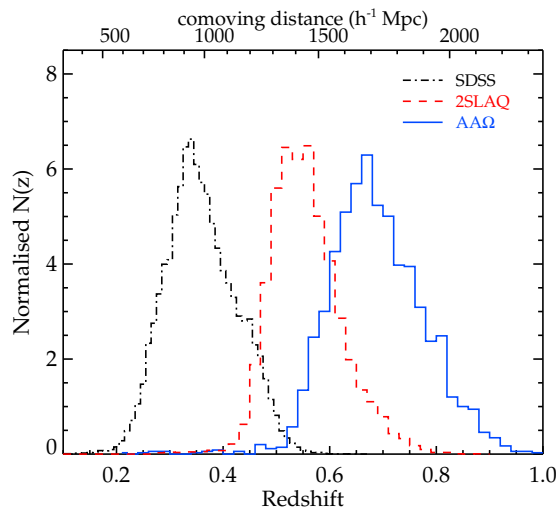}
\caption{Redshift distribution of the three LRG samples inferred from the 
redshift surveys used in their selections.}
\label{fig:nz}
\end{figure}

These simple methods have been proven to be highly effective in selecting the
intrinsically luminous early-type galaxies in the targeted redshift ranges as
demonstrated by the SDSS--LRG, 2SLAQ and AAT--AA$\Omega$ redshift surveys
\citep{Eisenstein01,Cannon06,Ross08}. Although the LRG photo-z in these 
redshift ranges can be estimated quite accurately \citep{Padmanabhan07,Collister07}, 
we decided to base our study on the colour--magnitude cuts because 
a well defined photo-z error distribution is needed for the deconvolution 
to recover the real redshift distribution and could bias the analyses 
of the results. The colour-magnitude cut techniques used in the above spectroscopic 
surveys, applied to the entire SDSS DR5 dataset (only Northern Galactic Cap), 
results in $\approx 1.5$ million LRG candidates and the redshift distribution of 
the survey is assumed for the corresponding photometric sample. The
outlines of the selection algorithms with the emphasis on any differences in our
criteria to that of the spectroscopic surveys are given below (readers are 
referred to \citealt{Eisenstein01,Cannon06,Ross08} for the detailed
descriptions of the selection criteria). 
The number-redshift relations, $N(z)$, (shown in Fig. \ref{fig:nz}) 
used in the model predictions have been calibrated to include those differences. 
The summary of the three LRG samples is given in Table \ref{tab:data}. 

For the following sections, all magnitudes and colours are given in SDSS $AB$
system (unless otherwise stated) and are corrected for
galactic extinction using the galactic dust map of \cite*{Schlegel98}.

\begin{table}
	\centering
	\caption{Summary of the LRG samples used in the cross-correlation analyses. }
		\begin{tabular}{lcrcc}
        \hline
        \hline
Sample	    & $ \zbar$  &Number   &	Sky density & Magnitude  \\
                &                   &  &   $(\rm deg^{-2})$& (AB) \\
          \hline
SDSS	 & 	0.35  &  106\,699 & $\approx$13	& $ 17.5  \le r < 19.5 $\\
2SLAQ 	 &	0.55  &  655\,775 & $\approx$85 & $ 17.5 < i < 19.8$\\
AA$\Omega$ &    0.68  &  800\,346 & $\approx$105 & $19.8 < i \le 20.5$ \\

	\hline
	\hline
		\end{tabular}
		
 \label{tab:data}
\end{table}

\subsubsection{SDSS LRG}
\label{sec:sdsslrg}
The low redshift (median $z \approx 0.35$) LRG candidates are selected on the basis
of their colours and magnitudes following the ongoing SDSS-LRG spectroscopic survey
\citep[E01 hereafter]{Eisenstein01} which will contain more than $100\,000$ spectra 
and cover over $1 h^{-3}~ \rm{Gpc}^{3}$ when completed.
The survey is designed to be approximately volume-limited up to $z \approx 0.4$.
The targets are selected using $g-r$ and $r-i$ colour cuts with the magnitude
limit, $r_{\rm petro} < 19.5$. Two sets of selection criteria (\textit{Cut I}
and \textit{Cut II} in E01) are used to extract LRGs in two different (but
slightly overlapped) regions of the $gri$ colour space, separated by the turn-over 
of the $gri$ colour track at $z \approx 0.4$.

In addition to the criteria of E01, we also apply restriction on the bright
limit in the $r$-band, i.e. $r_{\rm{petro}} \geq 17.5 $. This is mainly because
\textit{Cut I} is too permissive and allows under-luminous objects to enter the
sample below redshift 0.2 and by imposing the bright limit, we restrict the
sample to only $z \ga 0.2$. The choice of $r_{\rm{petro}} \geq 17.5 $ merits a
brief explanation. The redshift-dependent luminosity threshold is implemented by
one of the selection rules, $ r_{\rm{petro}} < 13.1+c_{\parallel}/0.3$ (Eq. 4 in
E01), where $c_{\parallel}\approx g-r$ at $z\approx0.2$ corresponds to $g-r \approx
1.3$ on the $gri$ colour-colour track. This has been empirically confirmed to
work sufficiently well using the spectroscopic sample, with only a few objects
having $r_{\rm{petro}} < 17.5$ at $z > 0.2$ and vice versa.    

The LRG sample described above is then extracted from the SDSS DR5 imaging
database using the SQL query by setting the flag PRIMTARGET to GALAXY\_RED. This
yields a catalogue of approximately $200\,000$ objects which after applying the
bright flux cut mentioned above, becomes 106\,699 objects with the sky
surface density of $\approx$ 13 $\rm{deg}^{-2}$. The average redshift of the LRG candidates 
as inferred from the spectroscopic sample of $\approx$ 60\,000 SDSS--LRG is $z=0.35 \pm 0.06$.

\subsubsection{2SLAQ LRG}

The 2dF--SDSS LRG and Quasar Survey (2SLAQ) is the spectroscopic follow-up of
the intermediate redshift ($z > 0.4$) LRGs from photometric data of the SDSS
survey using the 2-degree Field (2dF) instrument on the 3.9-m
Anglo--Australian Telescope (AAT). The survey was completed in 2006 and
contains approximately $13\,000$ spectroscopically confirmed LRGs with  $0.4 < z
< 0.8$ in two equatorial strips covering $\approx$ 180 $\rm{deg}^2$ \citep[and
references therein]{Cannon06}. The primary and secondary sample of the survey
(\textit{Sample 8} and \textit{9}, respectively) were selected using the SDSS
$g-r$ versus $r-i$ colours in conjunction with the `de Vaucouleurs' $i$-band
magnitude, $(17.5 < i_{\rm deV} < 19.8)$. The colour cuts are similar to the
\textit{Cut II} used by E01 which targets the objects lie above the turning
point of the early-type galaxy track in $gri$ colour space. The turning point
is caused by the 4000 \AA\ break moving into the $r$-band at $z \approx 0.4$,
making $r-i$ colour increase rapidly whereas $g-r$ remains nearly constant at
1.6--1.7 mag until $z \approx 0.7$.    

In order to qualify as 2SLAQ LRG candidates, objects are required to have
$d_{\perp} \ge 0.65$ for the primary sample and $ 0.55 \le d_{\perp}< 0.65$ for
the secondary sample, where $d_{\perp}=(r-i)-(g-r)/8.0$. We shall only use the
LRG candidates extracted following the primary sample cut, designed to target
higher redshift candidates than the secondary sample. We also utilise the
star--galaxy separation criterion used by the 2SLAQ survey which has been proven
to be very effective and the stellar contamination in the LRG sample is only 5
per\,cent. The primary sample contains 67 per\,cent of all 2SLAQ LRGs and has an
average redshift of $z=0.55 \pm 0.06$. Applying the primary target selection,
including the star-galaxy separation criteria, on the DR5 `best' imaging
database in the NGC, a sample of 655\,775 photometrically classified LRGs is
returned. Objects with BRIGHT or SATURATED or BLENDED but not DEBLENDED flags
are not included in our sample.

\subsubsection{AA$\Omega$ LRG}

Our new high redshift LRG sample is based on the AAT--AA$\Omega$ LRG Pilot run
\citep[and references therein]{Ross08}, using the 2dF instrument on the AAT. The
survey was carried out as a `Proof of Concept' for a large LRG redshift survey,
VST--AA$\Omega$ \textit{ATLAS} survey. It was designed to target potential 
LRGs out to $z \approx 1.0$ with the
average redshift of 0.7. Three different sets of selection criteria were
employed in selecting the targets in order to test the AA$\Omega$
spectrograph's ability to obtain reliable redshift with the minimum exposure
time in average conditions. They observed over $\approx$ 10 $\rm{deg}^2$ in three
2dF fields and the survey contains 1270 unique galaxy spectra with 804 high-confidence 
LRG redshifts.

The selection rules used here follow the colour-magnitude
cuts which utilise the $riz$ colour plane and the `de Vaucouleurs' $i$-band magnitude.
This selection forms the main part of the survey. In summary, the colour cuts exploit
the upturn of the early-type galaxy track similar to that used by 2SLAQ and SDSS
LRGs survey in selecting $z > 0.4$ LRGs with $gri$ colours. But in the $riz$
colour space, the upturn occurs between redshift 0.6 and 0.7 as the 4000 \AA\
feature moves into the $i$ band hence makes it ideal for selecting potential LRG
targets for the intended redshift range. The star--galaxy separation procedure uses
the $z$-band photometry, akin to the method which has proven effective
in the SDSS- and 2SLAQ-LRG redshift survey where a similar procedure were performed 
using the $i$-band photometry. Our star--galaxy separation algorithm only loses genuine 
LRGs at a sub-percent level and leaves $\approx$16 per\,cent stellar contamination in the sample.

The $riz$ selection has been proven to work reasonably well, resulting in  the 
sample having average redshift $z=0.68 \pm 0.07$. The
redshift distribution is further confirmed by the ongoing AAT--AA$\Omega$ LRG
project, a down-sized version of the VST--AA$\Omega$ \textit{ATLAS} survey,
designed to observed several thousands LRG redshifts for photo--z calibration
and clustering evolution study. The $N(z)$ used in the model prediction of ISW
signal comes from $\approx$ 2000 AA$\Omega$ LRG redshifts taken during a run in
June 2008 (Sawangwit et al. 2009, in prep.) as well as those from \cite{Ross08}.

The SDSS DR5 `best' imaging database contains 800\,346 photometric objects that
satisfied the AA$\Omega$ LRG selection rules including the necessary
star--galaxy separation performed in the $z$-band. As with the 2SLAQ LRG sample,
objects with BRIGHT or SATURATED or BLENDED but not DEBLENDED flags are
discarded from our sample.

\section{Theoretical prediction}   
\label{sec:theory}    

The secondary CMB anisotropy caused by the time-varying gravitional potential, 
$\Phi$, is known as the Integrated Sachs-Wolfe (ISW) effect. 
As the CMB photons traverse such regions, the temparature perturbation associated 
with the time dependent potential is given by  

\begin{equation}
\delta_{T}^{\rm{ISW}}(\bmath{\hat{n}}) \equiv \frac{\Delta_T^{\rm{ISW}}(\bmath{\hat{n}})}{T_0} 
=- 2 \int_{0}^{z_{{LS}}} dz ~\frac{1}{c^2} \frac{\partial \Phi}{\partial z}(\bmath{\hat{n}},z)
\label{equa:isw}
\end{equation}

\noindent where $\Phi$ is the Newtonian gravitational potential at redshift
$z$, $\bmath{\hat{n}}$ is a unit vector along a line of sight, $T_0=2.725 \,\rm{K}$ 
is the CMB temperature at present time and $z_{LS} \approx 1089$ is the redshift at 
the surface of last scattering. 

The gravitational potential, $\Phi$, is related to the matter density fluctuation via 
Poisson's equation (Eq. 7.14, \citealt{Peebles80});

\begin{equation}
\nabla^2 \Phi (\bmath{\hat{n}},z)= 4\pi G a^2 \rho_{\rm{m}}(z) \, \delta (\bmath{\hat{n}},z) 
\end{equation}

\noindent where $a$ is the scale factor normalised to unity at redshift zero. By recalling that 
$\rho_{\rm{crit}}(0)=3H_{0}^2/8\pi G$ and $\Omega_{\rm{m}}=\rho_{\rm{m}}(0)/\rho_{\rm{crit}}(0)$, 
the Fourier transform of the gravitational potential is

\begin{equation}
\Phi(\bmath{k},z)= -\frac{3}{2} \Omega_{\rm{m}} \left(\frac{H_0}{k}\right)^2 \frac{\delta(\bmath{k},z)}{a}.
\label{equa:Phik}
\end{equation}

Unfortunately, the ISW contribution to the CMB primary anisotropy is
less than 10 per\,cent for $l \ga 10$ and to make matters worse, the
total anisotropy signal is dominated by cosmic variance at smaller $l$ (i.e.
larger angle) where most of the ISW signal is expected to be
\citep[e.g.][]{Hu04}. To isolate the ISW signal one must cross-correlate the
temperature fluctuation with a tracer of gravitational potential projected on
the sky \citep{Crittenden96}. For this purpose, one can use the simple 2-point
statistics to compute the angular cross--correlation of the temperature
and galaxy fluctuation maps in real space, 

\begin{equation}
w_{gT}(\theta) = { \langle\delta_{g}(\bmath{\hat{n}_1}) \,\Delta_{T}(\bmath{\hat{n}_2})\rangle }
\end{equation}

\noindent where $\bmath{\hat{n}_1} \bmath{\cdot} \bmath{\hat{n}_2}=\cos{\theta}$. 
To calculate the theoretical expectation for the real space cross--correlation, 
we start by computing the angular cross--power spectrum of the galaxy overdensity and 
ISW temperature perturbation fields;

\begin{equation}
C_{gT}^{\rm{ISW}}(l) \equiv \langle \delta_{g,lm} \, \Delta_{T,l'm'}^{\ast} \rangle.
\label{equa:clfull}
\end{equation}

Firstly, we need to expand the galaxy density fields, $\delta_g(\bmath{\hat{n}},z)$, 
in spherical harmonics and Fourier transform them. For a galaxy survey with a selection 
function $\phi_g(z)$ and linear bias $b_g(z)$, this is 
 
\begin{eqnarray}
\delta_{g,lm}&=&i^l \int \frac{d^3k}{(2\pi)^3} \int dz\,4\pi j_l(k\chi) \, Y^{\ast}_{lm}(\bmath{\hat{k}})  \nonumber \\
             & &   ~~~~~~~~~~~~~~~~~~~~ \times  b_g(z) \, \phi_g(z) \, \delta(\bmath{k},z)
\label{equa:glm}  
\end{eqnarray}

\noindent where $j_l(y)$ is the spherical Bessel function of the first kind of rank 
$l$, $Y_{lm}(\bmath{\hat{k}})$ is the spherical harmonic function and $\chi$ 
is a comoving distance which is an implicit function of $z$ through the relation $d\chi=c/H(z) \, dz$.
In obtaining Eq. \ref{equa:glm}, we use the orthonormality of $Y_{lm}$ in their 
expansion of a plane wave (e.g. \citealt{Scharf92});

\begin{equation}
{\rm{exp}}({-i\bmath{k}\bmath{\cdot}\bmath{\hat{n}}\chi})=4\pi \sum_{lm} i^{l} \, j_l(k\chi) \, Y_{lm}(\bmath{\hat{n}}) \, Y_{lm}^{\ast}(\bmath{\hat{k}})
\label{equa:plane} 
\end{equation}

Similarly, for the ISW temperature fluctuation, by putting together
Eq. \ref{equa:isw}, \ref{equa:Phik} and \ref{equa:plane}, this is

\begin{eqnarray}
\Delta_{T,lm}^{\rm{ISW}}&=&i^l \int \frac{d^3k}{(2\pi)^3} \int dz\,4\pi j_l(k\chi(z)) \, Y^{\ast}_{lm}(\bmath{\hat{k}})  \nonumber \\
             & &   ~~~~~~~~~~~\times 3 \Omega_{\rm{m}} T_{0} \left(\frac{H_0}{kc}\right)^2 \frac{\partial}{\partial z} \left[\frac{\delta(\bmath{k},z)}{a(z)}\right]
\label{equa:Tlm}   
\end{eqnarray}

For a flat--sky approximation \citep{Limber53}, following \cite{Afshordi04a} 
and realising that in linear perturbation theory 
$\delta(\bmath{k},z) = D(z) \, \delta(\bmath{k},0)$ and

\begin{equation}
\langle \delta(\bmath{k_1}) \, \delta(\bmath{k_2}) \rangle = (2\pi)^3 \, \delta_{\rm{Dirac}}(\bmath{k_1}-\bmath{k_2}) \, P(k) \, ,
\end{equation}
from Eq. \ref{equa:clfull}, \ref{equa:glm} and \ref{equa:Tlm}
, $C_{gT}^{\rm{ISW}}(l)$ can be simplified to

\begin{equation}
C_{gT}^{\rm{ISW}}(l) = \frac{4}{(2l+1)^2} \int dz \, P(k) \, W_{\rm{ISW}}(z) \, W_g(z).
\end{equation}

\noindent $W_{\rm{ISW}}(z)$ and $W_g(z)$ are the ISW and galaxy window functions defined as

\begin{equation}
W_{\rm{ISW}}(z) \equiv 3 \Omm T_0 \left(\frac{H_0}{c}\right)^2  \frac{d}{dz} \left[\frac{D(z)}{a(z)}\right] 
\end{equation}

\noindent and

\begin{equation}
W_g(z) \equiv b_g(z) \, \phi_g(z) \, D(z) 
\end{equation}

\noindent where $k \approx {(l+1/2)/\chi(z)}$, $D(z)$ is the linear growth factor given by the 
fitting formula of \citet{Carroll92} and $P(k)$ is the linear power spectrum 
at redshift zero. The survey selection function is given by 

\begin{equation}
\phi_{g}(z) \equiv \frac{\chi^2 n_c(\chi)}{\int d\chi \, \chi^2 n_c(\chi)}=n(z) \frac{H(z)}{c} 
\end{equation}

\noindent where $n_c(\chi)$ is the comoving number density and $n(z)$ is the normalised 
redshift distribution, $N(z)$, of the galaxies in the survey. Finally, 
$w_{gT}^{\rm{ISW}}(\theta)$ is related to the cross--power spectrum via the 
expansion in Legendre polynomials;

\begin{equation}
w^{\rm{ISW}}_{gT}(\theta) = \sum_{l=2}^{\infty} \frac{2l+1}{4\pi}\, P_{l}(\cos\theta)\, C_{gT}^{\rm{ISW}}(l).
\label{equa:wgt}
\end{equation}

We set the monopole ($l=0$) and dipole ($l=1$) contribution to zero, 
as it is done in the \textit{WMAP} maps (\S \ref{sec:wmap5}). The contributions
of the monopole and dipole are significant and overpredict $w^{\rm{ISW}}_{gT}$ 
by $\approx$10 per\,cent \citep{Cabre06}. 
The summation in Eq. \ref{equa:wgt} converges earlier than 
$l \approx 500$ but we set our upper limit to $l=1000$ which 
provides sufficiently stable models without sacrificing too much computing time. 
The linear power spectrum is computed using

\begin{equation}
P(k)= A~k^{n_s}~T^2(k)
\end{equation}

\noindent where $n_s$ is the scalar spectral index and $A$ is the normalisation factor 
with the value set by $\sigma_8$. We use the transfer function, $T(k)$, 
fitting formula of \cite{Eisenstein98}. Our fiducial models assume a 
$\Lambda$CDM Universe with $\Omega_{\Lambda}=0.73,~\Omega_{\rm{m}}=0.27,
~f_{\rm{baryon}}=0.167,~\sigma_{8}=0.8$, $h=0.7$ and   
$n_s=0.95$. Note that, for a flat Universe with 
$\Omm=1$, the linear growth factor is equal to the scale factor, $a$, at all redshifts 
and $W_{\rm{ISW}}(z)$ vanishes, hence in this case we expect no correlation 
between large-scale-structure and the CMB. 
 
\begin{figure*}
\vspace{0.8cm} 
\centering
 \includegraphics[scale=0.55]{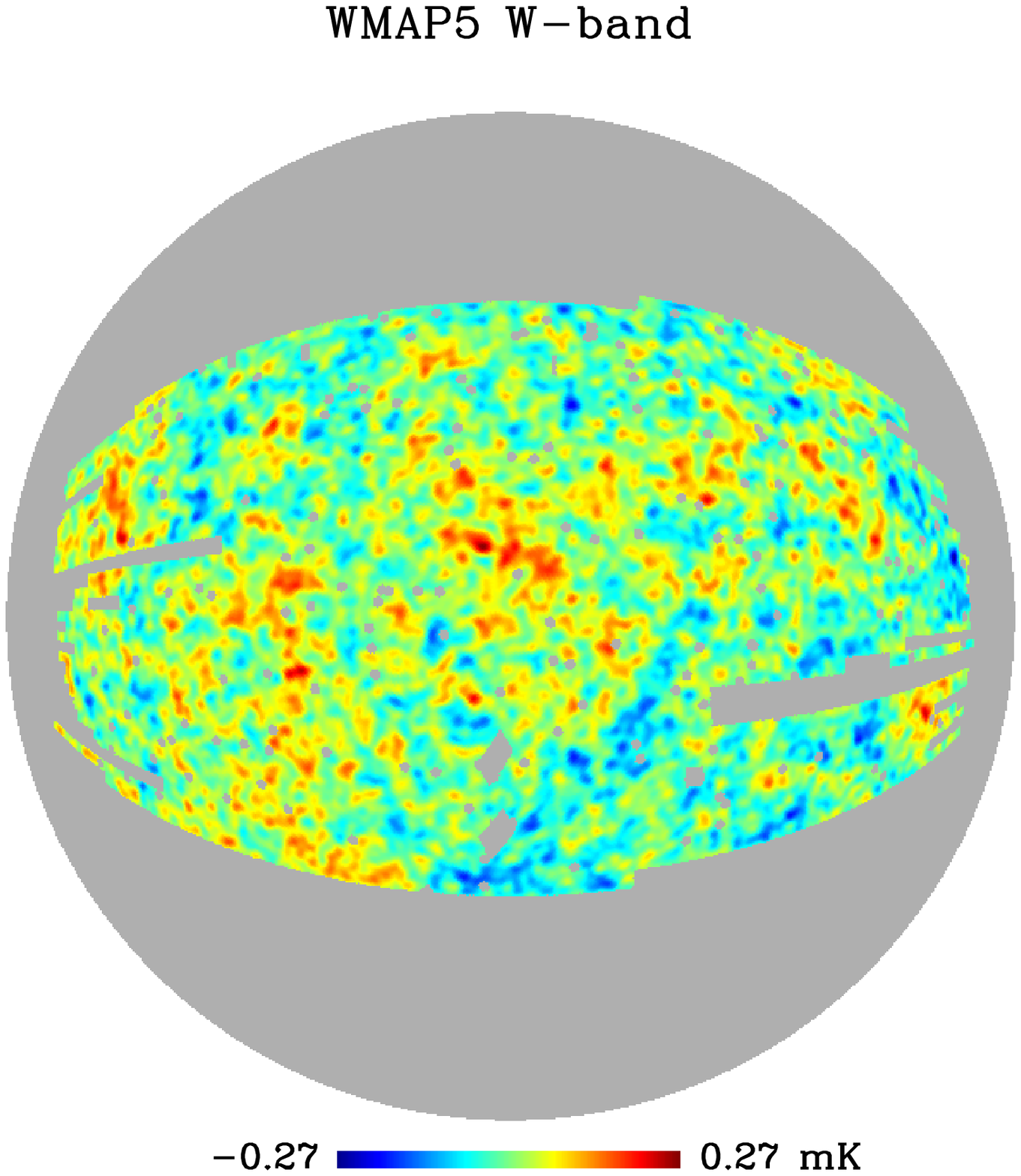}\includegraphics[scale=0.55]{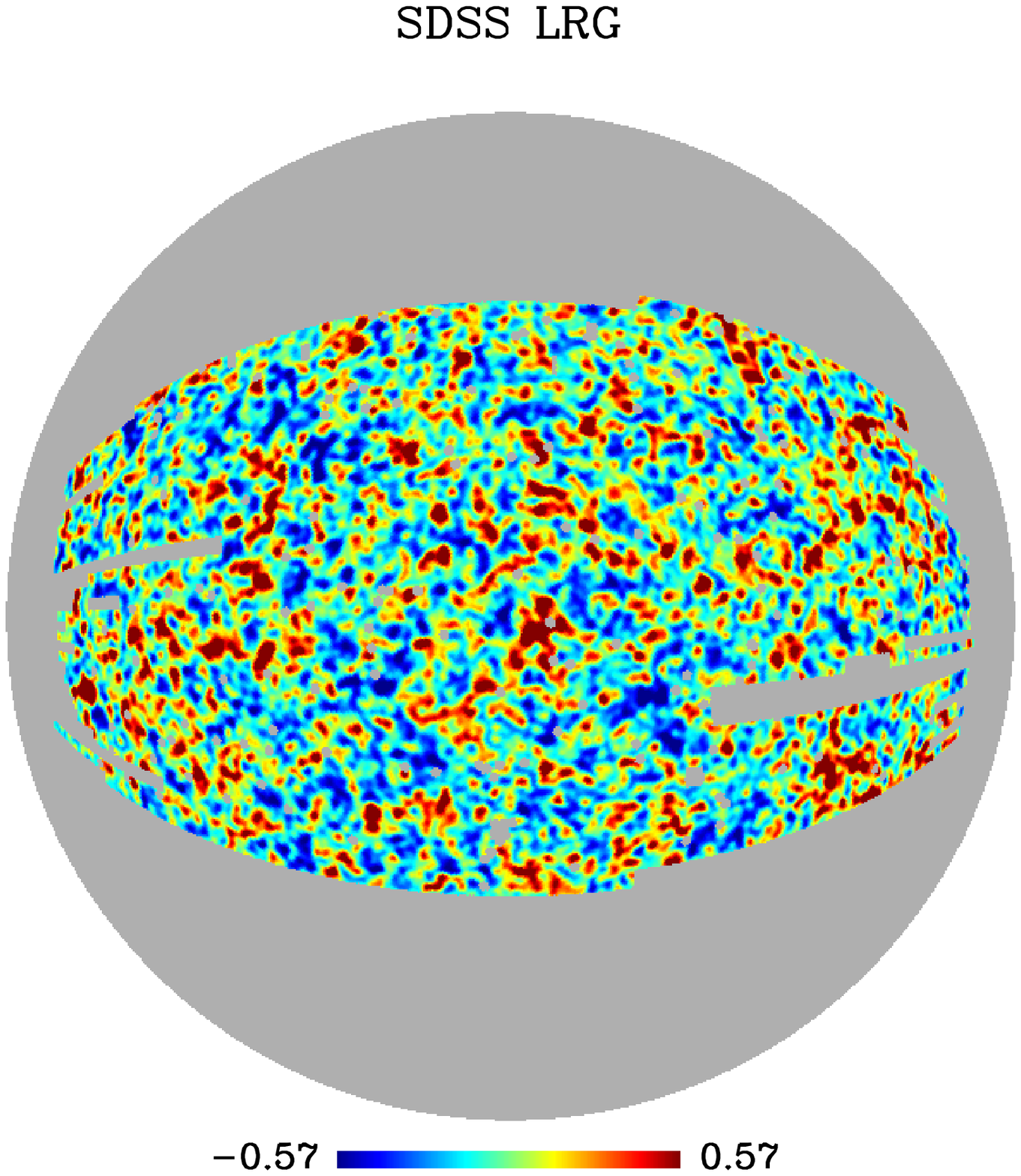} \includegraphics[scale=0.55]{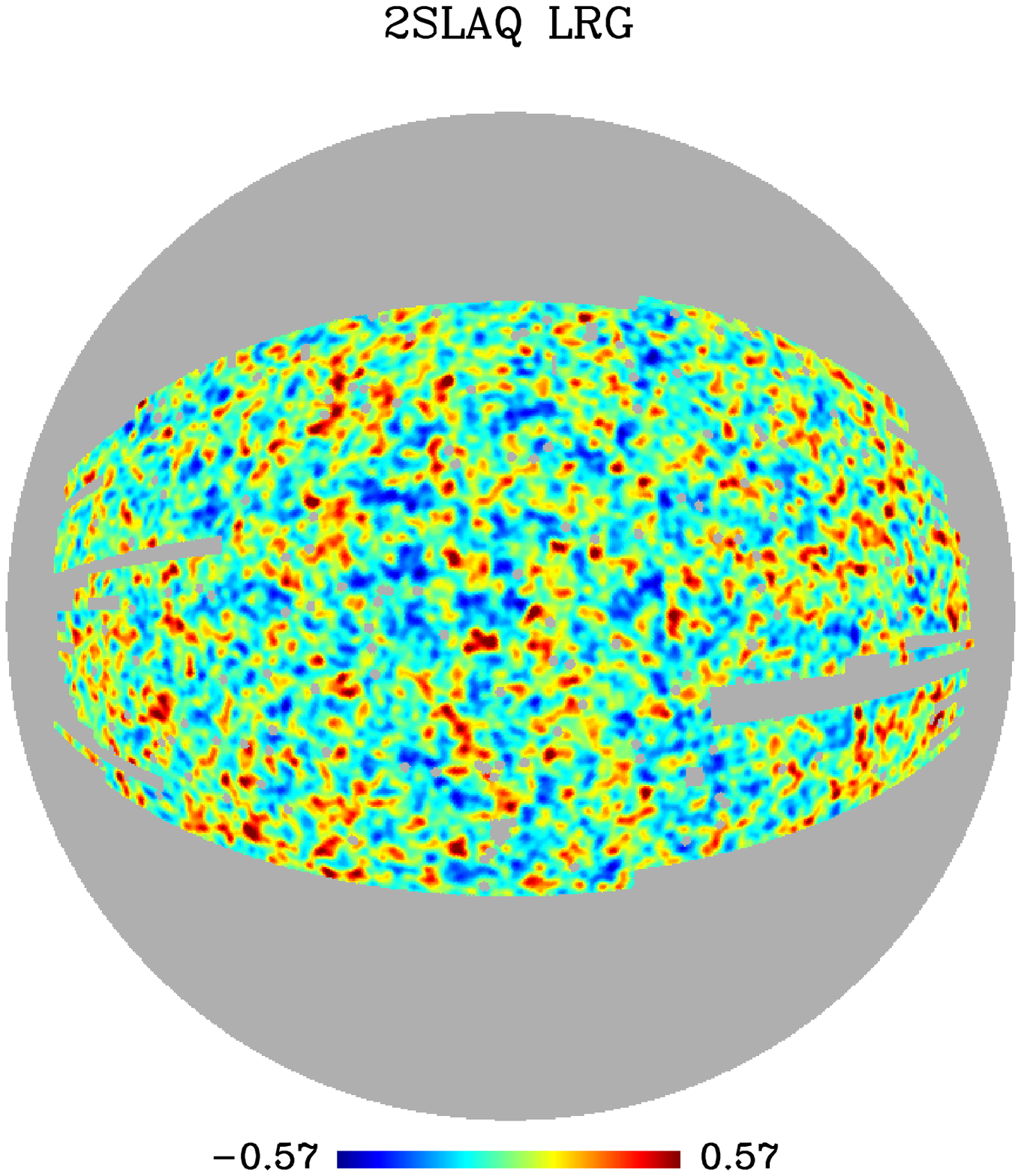}\includegraphics[scale=0.55]{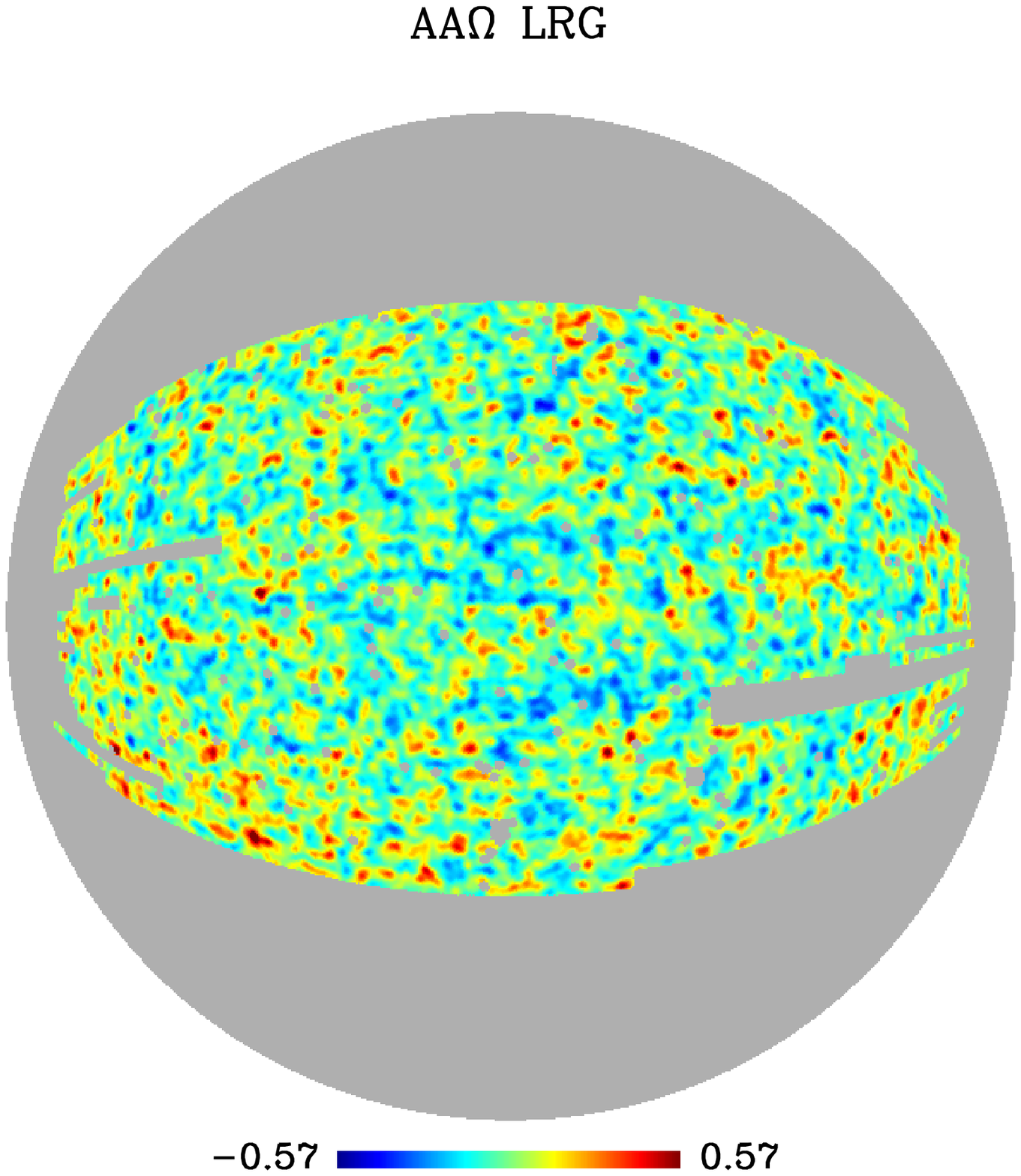}

\caption{The 1$\deg$ smoothed map of W-band data and galaxy number overdensity for SDSS, 2SLAQ and AA$\Omega$ LRG (Ubercal) after applying KQ75 and SDSS-DR5 mask.}
	\label{map}
\end{figure*}

\section{Cross-correlation technique}
\label{sec:technique}

Firstly, each galaxy sample is pixelised into equal area pixels on the sphere 
using the HEALPix \citep{HEALPixref} format, following the standard
resolution and ordering scheme of the publicly-available \textit{WMAP}5 
temperature map (i.e. nested, res=9). The most conservative temperature
mask, extended temperature analysis (KQ75), plus point source catalogue mask are
then applied to the temperature maps (\S \ref{sec:wmap5}) and the pixelised galaxy distributions,
discarding approximately 30 per\,cent of the entire sky. Additionally, in order
to estimate fairly the galaxy background density and a robust cross--correlation
result, the DR5 coverage mask including quality holes are applied to the data. We
only restrict the data to the most contiguous region of the NGC and therefore
exclude the SDSS stripes 39, 42 and 43 in the DR5 coverage mask. After applying `KQ75 $\cup$ point
source $\cup$ DR5' mask, 516,507 out of 3,145,728 pixels (all sky) are admitted for
the cross--correlation analysis. 

The galaxy number overdensity, $\delta_g(\bmath{\hat{n}})$, is then calculated from the
galaxy distribution maps and assigned to each pixel;
 
\begin{equation}
\delta_g(\bmath{\hat{n}})=\frac{n_g(\bmath{\hat{n}})-\overline{n}_g}{\overline{n}_g}
\end{equation}

\noindent where $n_g$ and $\overline{n}_g$ are the number of galaxies and its average for 
the sample of interest. Fig. \ref{map} shows the W-band temperature fluctuation map and 
$\delta_g$ map for SDSS, 2SLAQ and AA$\Omega$ LRG, smoothed with Gaussian beam of 1$\deg$ Full 
Width at Half Maximum (FWHM). 

The two--point cross--correlation function at angular separation $\theta$ is computed using

\begin{equation}
w_{gT} (\theta)= \frac{\sum_{ij} f_i\delta_g(\bmath{\hat{n}_i}) \,f_j \Delta_T(\bmath{\hat{n}_j})}{\sum_{ij}f_i \, f_j}
\label{equa:estwgt}
\end{equation}

\noindent where $f_i$ is the fraction of pixel $i$ within the unmasked region,
$\bmath{\hat{n}_i}\bmath{\cdot}\bmath{\hat{n}_j}=\cos{\theta}$ and $\Delta_{T}$
is the CMB temperature anisotropy measured by \textit{WMAP}5 with the monopole
and dipole contribution subtracted off. However, as we use relatively fine
resolution pixels and weighting by the unmasked fraction does not alter our
measurement, Eq. \ref{equa:estwgt} is simply $w_{gT}(\theta)=\langle
\delta_{g}(\bmath{\hat{n}_1}) \, \Delta_{T}(\bmath{\hat{n}_2})\rangle$.

It is a well known fact that bins in the correlation function are correlated
because the same points (or pixels in this case) can appear in many different
pairs which are included in different bins, especially at large scales. To
correctly estimate the statistical significance of the results, one needs to
consider the full covariance matrix, \mathbfss{C$_{ij}$}. Here, we construct the
full covariance matrices using the \textit{jackknife} re-sampling. In order to
obtain a sufficiently stable covariance matrix, the jackknife subsamples of
approximately twice the number of angular bins being considered are needed. For
the number of angular bins considered in this study, we split the masked
temperature/overdensity map into 24 subfields with approximately equal area. The
24 jackknife subsamples are constructed from these fields, each one leaving out
a different subfield. The $w_{gT}(\theta)$ are computed for each jackknife
subsample and the covariance matrix is 
  
\begin{eqnarray}
\mathbfss{C}_{ij} &=& \frac{N_J-1}{N_J}\sum_{m=1}^{N_J} [ (w_{gT,m}(\theta_{i})-\overline{w_{gT}}(\theta_{i}))  \nonumber \\
& & ~~~~~~~~~~~~~~~~~\times(w_{gT,m}(\theta_{j})-\overline{w_{gT}}(\theta_{j})) ]    
\label{equa:covariance}
\end{eqnarray}

\noindent where $N_J=24$ in this case, ${w_{gT,m}}(\theta_{i})$ and
$\overline{w_{gT}}(\theta_{i})$ are the cross--correlation measured from the
$m$th jackknife subsample and the average of all the subsamples in the $i$th
bin, respectively. Note that the difference between $\overline{w_{gT}}(\theta)$
and $w_{gT}(\theta)$ estimated using the whole sample is negligible. The reason
for multiplying $N_J-1$ is because the jackknife subsamples are not independent.
The statistical uncertainty for each individual angular bin is contained in the 
diagonal elements of the covariance matrix.

\begin{figure*}
	
\includegraphics[scale=0.5]{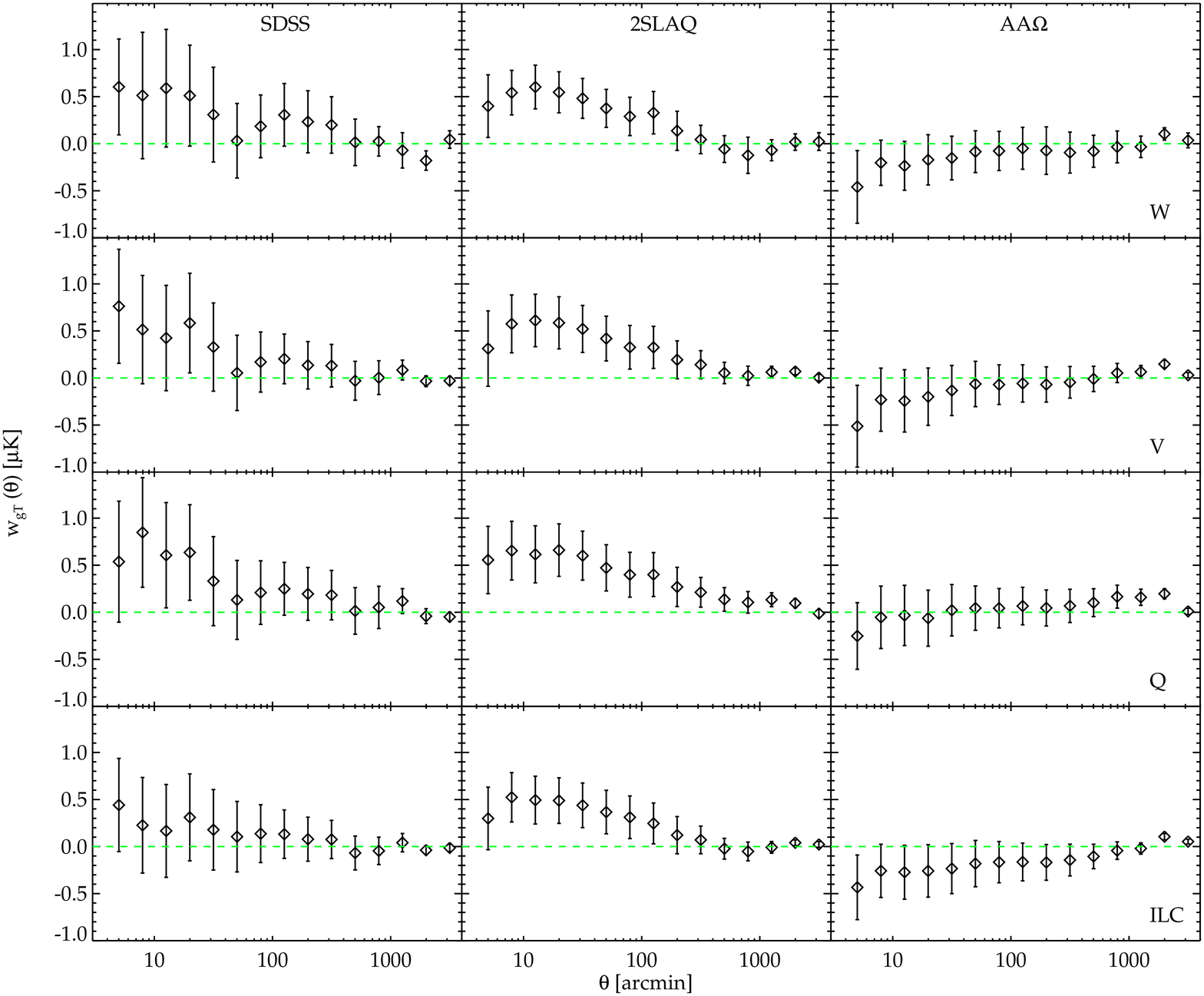}
\caption{The cross-correlation results of \WMAP5 ~\W, \V ~and \Q ~band including
the ILC map (top to bottom) with the SDSS, 2SLAQ and AA$\Omega$ LRG (left to
right).}
\label{fig:band}
\end{figure*}

\begin{figure*}
\hspace{-0.4cm}
	\centering$
\begin{array}{ll}
 
\hspace{-0.5cm} \includegraphics[scale=0.35]{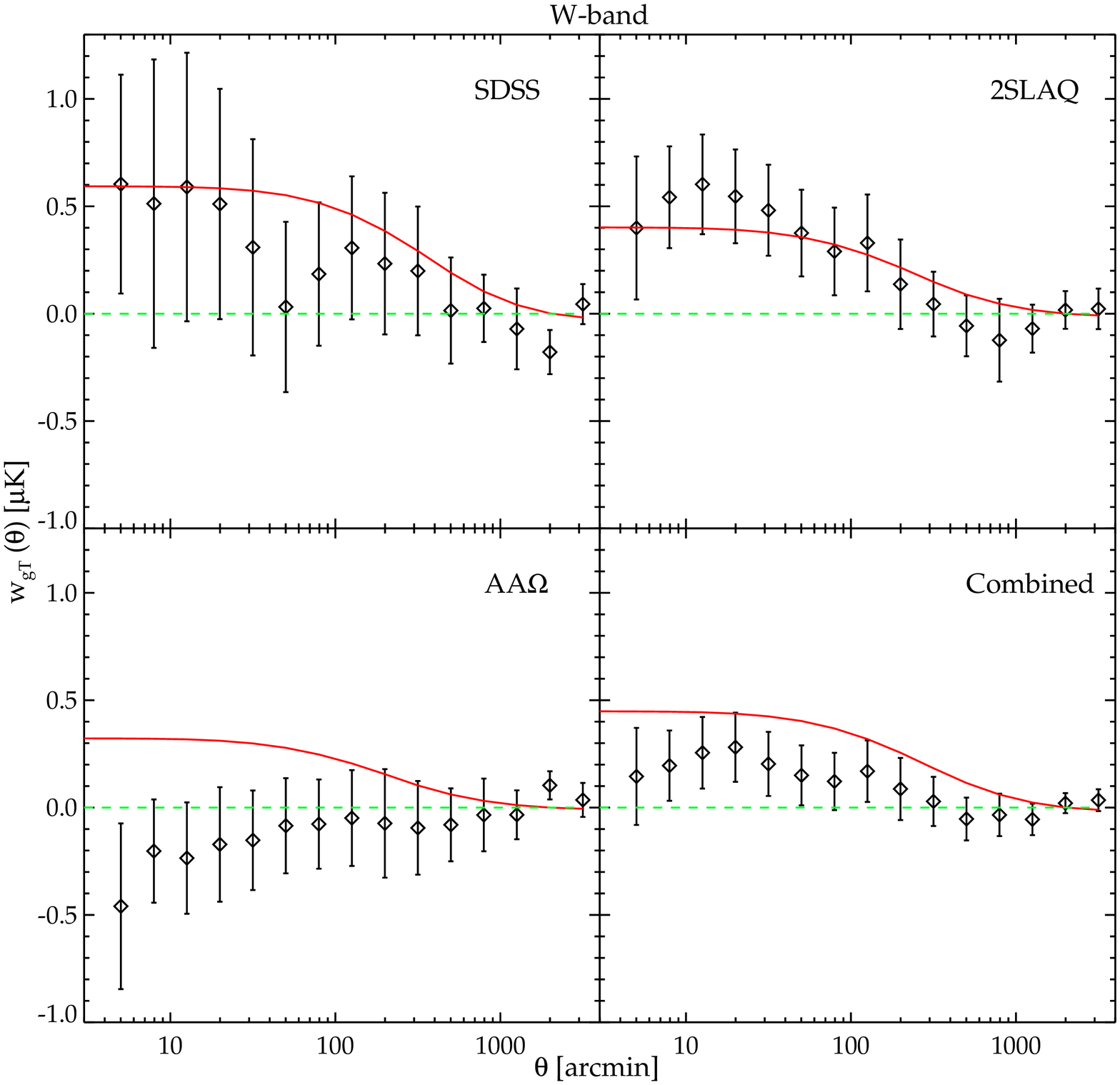}&
 \hspace{-0.8cm} \includegraphics[scale=0.35]{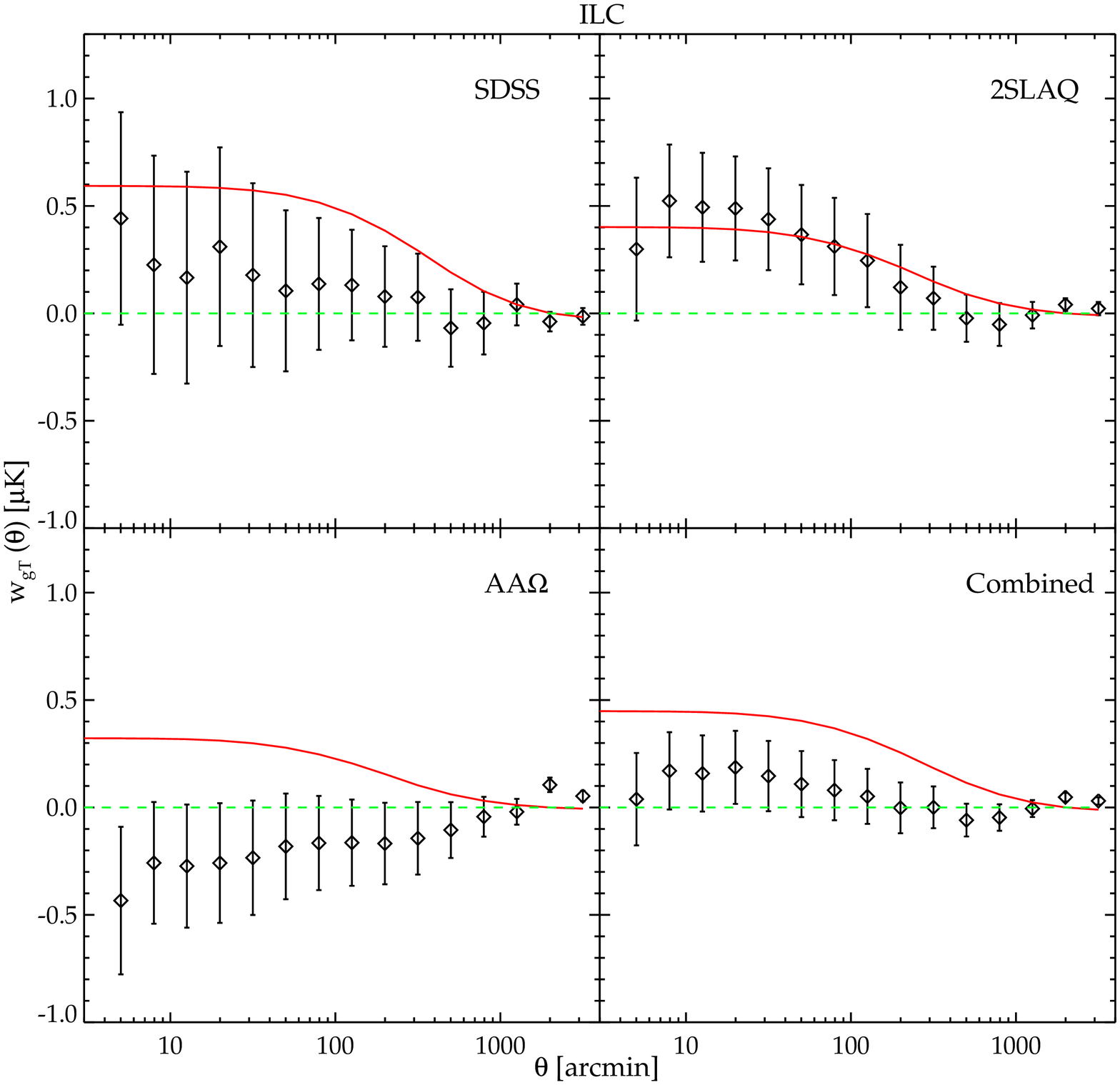}
\end{array}$
\caption{The LRG--\textit{WMAP5} cross-correlation results using \textit{W}-band
and ILC map compared to the theoretical predictions (red solid lines), assuming
the standard $\Lambda$CDM and the galaxy linear bias ($b_{g}$) of 2.10, 1.99,
2.2 and 2.1 for SDSS, 2SLAQ, AA$\Omega$-LRG and the combined sample
respectively. The stellar contamination correction for each sample has been
applied to the corresponding model. In the `Combined' panels, the
cross--correlation results of the quadrature--error weighted mean of the three
LRG samples are compared to the average model
predictions.} \label{fig:mod}
\end{figure*}

\section{Results and Analysis}
\label{sec:result}
\subsection{LRG--\textit{WMAP}5}

The cross--correlation results of the LRG distributions with the \textit{WMAP}5
temperature maps using the three highest-frequency data plus the ILC are shown
in Fig. \ref{fig:band}. The errors are $1\sigma$ statistical errors estimated
from jackknife re-sampling of 24 subfields as described in \S
\ref{sec:technique}. Generally, the results using different \textit{WMAP} bands
are in good agreement (within the $1\sigma$ error) for all three LRG samples. The
achromatic results indicate that the contamination from effects such as dust,
synchrotron and free-free emission which are frequency-dependent in nature are
minimal or at least sub-dominant compared to our statistical uncertainties. This
also applies to a lesser extent to the thermal Sunyaev--Zeldovich
\citep{Sunyaev80} effect, although for the bands shown, the difference in the SZ
and CMB  spectral slopes is only $\approx$30 per\,cent. However, we shall see in
\S \ref{sec:rotate} that there is still a strong suggestion that other
systematic effects may still be contaminating the SDSS and 2SLAQ results.

We first consider our new and higher redshift sample of 800\,000 AA$\Omega$ LRGs.
This sample shows virtually no positive correlation with the CMB data. If
anything, the data show a slight anticorrelation out to large scales, possibly to
$\theta \la 1\deg$ ($\approx 30 h^{-1}$Mpc at the median redshift of the sample),
although the signal to noise is still low. This weak anticorrelation is observed
in all \textit{WMAP}5 frequency bands under study here (the most right column of
Fig. \ref{fig:band}) with the exception of the \Q band which only shows zero
correlation at best with a possible zeropoint shift towards 
very large scales. As for the SDSS and 2SLAQ results the cross--correlation
with the ILC map gives a systematically lower amplitude (more negative in
AA$\Omega$ case) than other bands. Given the relatively large scales of the null
result in the AA$\Omega$--\textit{WMAP}5 CCF and the amplitude of the expected
ISW signal (see Fig. \ref{fig:mod}), it would seem extremely unlikely that
the positive correlation of the ISW effect could be cancelled out by the negative
contribution from the thermal SZ effect. If this result is real and not due to
some systematic effects, the implications for the view that the Universal
expansion is accelerating, could be profound.  

In the case of the SDSS and 2SLAQ LRG samples, our results are similar
to those of the previous authors who have analysed similar datasets. We observe marginally 
significant positive correlations in the \textit{Q}, \V
and \W bands where the measured $w_{gT}(\theta)$'s are  similar in terms of their
amplitudes and angular extents for each sample although the signal is weaker
in the SDSS sample. The ILC results are slightly lower than the other bands in
both samples but otherwise still within $1\sigma$ error. Our SDSS results can be
compared to the lowest redshift--bin sample of \cite{Scranton03} who used the
LRGs extracted from the SDSS DR2 following \cite{Eisenstein01} but with a much
fainter magnitude limit, $i < 21$, and divided their samples into redshift
slices using photo--z. The results are similar in terms of amplitude but our
errors are slightly smaller due to our larger area coverage ($\approx$7600 deg$^2$
as opposed to $\approx$3400 deg$^2$) although their object numbers are $\approx$7
times higher than ours owing to the broader $N(z)$ and fainter flux cut. The
2SLAQ results are comparable to the `SDSS LRG' results of \cite{Giannantonio08}.
These authors used the MegaZ-LRG photo-z catalogue of \cite{Collister07},
covering the redshift range 0.4--0.7  with a colour-magnitude selection similar to our
2SLAQ sample but a slightly fainter flux limit, $i_{\rm{deV}} < 20$ as opposed to
19.8. In the LRG panel of their Fig. 4, we see that their result has similar
amplitude and errors (jackknife) to our 2SLAQ result. Although their Monte
Carlo methods give somewhat larger errors than the jackknife estimations, the statistical
significance estimated using errors drawn from both methods are very similar,
2.2--2.5$\sigma$ for their LRG catalogue. \citet{Padmanabhan05b} has also
performed the analysis with a similar LRG sample but using the angular
cross--power spectrum, $C_{l}$, making a direct comparison to our results
difficult. The sample these authors used is somewhat similar to the
\cite{Eisenstein01} selection but with the flux cut as faint as 2SLAQ in
`CutII', resulting in a redshift distribution similar to our SDSS and 2SLAQ
LRG samples combined, although they limited the redshift of the sample to $ 0.2 < z <
0.6$ using their template-fitting photo-z. The positive correlation is detected
at 2.5$\sigma$, similar to \cite{Giannantonio08} although the sample they used only
covers half as much  sky. We conclude that our analyses are broadly
reproducing previous results in these  $ 0.25 < z < 0.6$ LRG redshift ranges,
both in terms of their amplitude and statistical significance.

 \begin{figure}
 \hspace{-0.5cm}
 	\centering
  \includegraphics[scale=0.46]{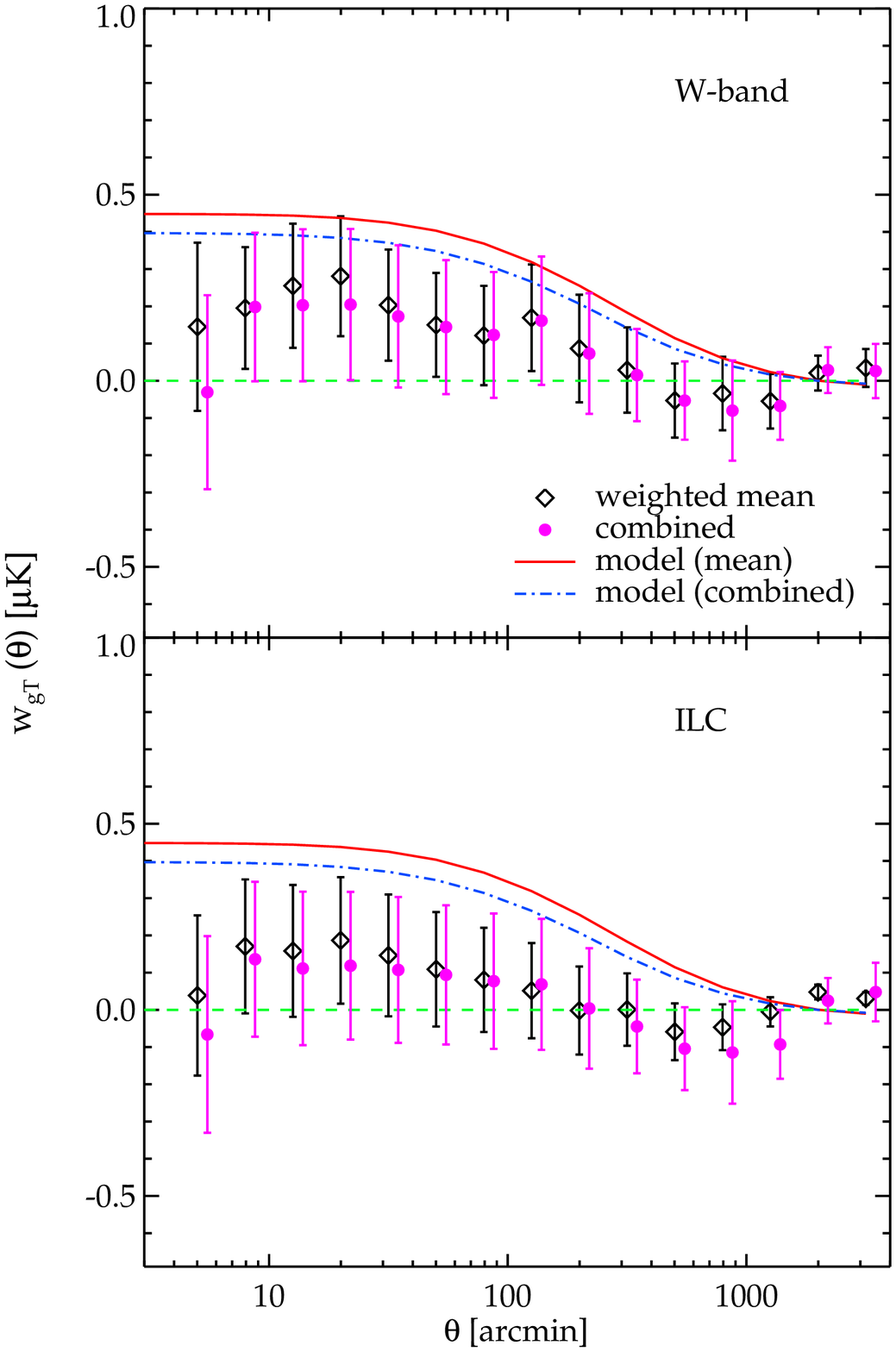}
     \caption{\textit{Top}: The W-band cross--correlation results of the
     combined sample (solid circles) compared to the quadrature--error weighted
     mean of the three LRG sample (diamonds). Also shown are the standard model
     predictions by taking a weighted mean (solid line) of the models of three
     LRG samples and for the combined sample (dot-dash line). \textit{Bottom}:
     Same as above but for the ILC map rather than \textit{W}-band data}
 	\label{fig:combined}
 \end{figure}

\begin{table*}
	\centering
	\caption{The significance tests of the cross--correlation results using the
	\textit{WMAP} \textit{W}-band data and ILC maps. The measurements are tested
	against the expected ISW prediction in the standard $\Lambda$CDM model and null result
	hypothesis. Column 5 gives the amplitudes and $1\sigma$ jackknife errors of
	the data binning between 12\arcmin--120\arcmin. Column 6 gives the
	significance of the deviation of the value in column 5 relative to ISW/null
	signal hypothesis.}
		\begin{tabular}{rlcrccc}
        \hline
        \hline
 \multicolumn{2}{c} {Sample} & $\zbar$&Number&$b_g(\bar{z})$ &$w_{gT}(12\arcmin-120\arcmin)$& Deviation significance  \\
 &      &       &        &              &  $\mu$K                    &   (ISW,null)  \\
          \hline
$W$-band:&SDSS	 & 	0.35  &  106,699 &   $2.10\pm0.04 $ & $0.25\pm 0.33$    &  $(1.0\sigma,0.8\sigma)$ \\
       &2SLAQ 	 &	0.55  &  655,775 &   $1.99\pm0.02 $ & $0.34\pm 0.21$  &  $(0.2\sigma,1.6\sigma)$  \\
       &AA$\Omega$ &    0.68  &  800,346 &   $2.20\pm0.02 $ & $-0.07\pm 0.20$ &  $(1.9\sigma,0.4\sigma)$  \\
       &AA$\Omega^{\ast}$ & 0.67 & 375,056 & $2.37\pm0.03 $ & $-0.10\pm 0.20$ &  $(2.2\sigma,0.5\sigma)$  \\
       &Combined  &  0.60     & 1,562,820&   $2.10\pm0.01 $ & $0.15\pm 0.17$  &  $(1.0\sigma,0.9\sigma)$   \\
       &Weighted mean&--&\multicolumn{1}{c}{--}& -- & $0.14\pm 0.14$ &  $(1.3\sigma,1.0\sigma)$    \\
       \hline
ILC map:&SDSS	 & 	0.35  &  106,699 &   $2.10\pm0.04 $ & $0.19\pm 0.33$    &  $(1.2\sigma,0.6\sigma)$  \\
       &2SLAQ 	 &	0.55  &  655,775 &   $1.99\pm0.02 $ & $0.27\pm 0.22$  &  $(0.5\sigma,1.2\sigma)$  \\
       &AA$\Omega$ & 0.68     &  800,346 &   $2.20\pm0.02 $ & $-0.18\pm 0.22$ &  $(2.2\sigma,0.8\sigma)$  \\
       &AA$\Omega^{\ast}$ & 0.67 & 375,056 & $2.37\pm0.03 $ & $-0.20\pm 0.21$ &  $(2.5\sigma,1.0\sigma)$  \\
       &Combined  &  0.60     & 1,562,820&   $2.10\pm0.01 $ & $0.07\pm 0.17$  &  $(1.4\sigma,0.4\sigma)$    \\
       &Weighted mean&--&\multicolumn{1}{c}{--}& -- & $0.07\pm 0.13$ &  $(2.0\sigma,0.5\sigma)$     \\
        \hline
	\hline
		\end{tabular}
		
 \label{tab:significance}
\end{table*}

\subsection{Comparison to Models}
\label{sec:comparison}
Fig. \ref{fig:mod} shows the comparison of our results to the theoretical
expectation as described in \S \ref{sec:theory}. The galaxy selection
functions used in construction of these models are given by the normalised
$N(z)$ of the sample as shown in Fig. \ref{fig:nz} (see also \S
\ref{sec:lrg}). The galaxy bias in the model is estimated from the angular
autocorrelation function, $w_g(\theta,\bar{z})$, of each LRG sample relative to
the underlying dark matter clustering,
$b_g^2(\bar{z})=\xi_{g}(r,\bar{z})/\xi_{\rm{m}}(r,\bar{z})$, where we assume the
linear scale-independent bias and measure its value at large scales ($\approx$10
$h^{-1}$Mpc). The matter $\xi(r,\bar{z})$ is estimated for the same fiducial
cosmology as described in \S \ref{sec:theory} and then projected onto the
sky using the corresponding $n_g(z)$. This gives an unbiased prediction which
can be compared to the measured $w_g(\theta,\bar{z})$ and allows $b_g(\bar{z})$
to be extracted from their relative amplitudes (see Sawangwit et al. in prep for
the full detailed analyses). Note that, we assume non-evolving bias and denote 
the bias estimated from each sample as the bias at the corresponding average 
redshift which is reasonable, given the narrow redshift ranges of our samples. 
The galaxy bias measured in this way can also be viewed as an effective value 
for each sample. The models shown in Fig. \ref{fig:mod} use $b_g(\bar{z})$ 
of $2.10\pm0.04$, $1.99\pm0.02$ and $2.20\pm0.02$ for the SDSS, 2SLAQ and AA$\Omega$samples, respectively.
These values are taken from Sawangwit et al. (in prep) and are compatible with the 
values measured by other authors, e.g. \citet{Tegmark06}, \citet{Padmanabhan07} 
whose $b_g(0.35)=1.9\pm0.07$ and $b_g(0.55)=1.85\pm0.05$ as compared to our SDSS 
and 2SLAQ LRG, respectively. The $b_g$ value of \citet{Tegmark06} was measured from 
a sample of $z\approx0.35$ LRGs similar to what we call SDSS LRG sample here but without
the bright limit cut (see \S \ref{sec:sdsslrg}) hence allows under--luminous objects and 
main galaxies into their sample. And as a result we expect their bias to be somewhat lower than ours.  

As emphasised earlier, the AA$\Omega$ LRG sample shows no positive correlation
with the \textit{WMAP}5 data and perhaps even a slight negative correlation. We
then combined the $W$-band data between 12\arcmin--120\arcmin, and found the
amplitude of the CCF and its jackknife error ($1\sigma$) is $-0.07\pm 0.2$
$\mu$K. This is consistent with the null hypothesis (only $\approx 0.4\sigma$
deviation) and rejects the ISW signal expected in the standard models at $\approx
1.9\sigma$  or at 5\% significance  after the stellar contamination has
been taken into account in the predicted signal (see \S \ref{sec:stellar}).
Performing a similar statistical analysis on the cross--correlation results
using the ILC map gives a slightly higher significance of rejecting the standard
model ISW hypothesis ($2.2\sigma$, see Table \ref{tab:significance}).

Additionally, to test that the zero correlation in the AA$\Omega$ results is not
due to its faint limit making the sample incomplete, we have cut the faint
limit of the sample back in steps of 0.25 mag to 20.0 (see \S
\ref{sec:photo} and Sawangwit et al. 2009). The amplitude of the CCF between
12\arcmin--120\arcmin~for $i < 20.25$ (denoted by AA$\Omega^{\ast}$ in Table
\ref{tab:significance}) is $-0.1\pm 0.2$ for $W$-band data and $-0.2\pm 0.21$
for the ILC map. The ISW model prediction is then re-computed taking into account the
corresponding $n(z)$ and linear bias, including the correction for stellar
contamination at the same level as the main AA$\Omega$ sample. The significance
of rejection of  the standard model for the $i < 20.25$ AA$\Omega$ sample is slightly
higher than that of the main AA$\Omega$ sample, at $2.2\sigma$ and $2.5\sigma$
for $W$-band and ILC map, respectively.

The measured $w_{gT}$ for the SDSS LRG agrees reasonably well with the theoretical
expectation at angles $\la 30 \arcmin$ although not at high statistical
significance. However, the same cannot be said for the angle beyond this scale
and up to $\approx 600 \arcmin$ where the cross--correlation appears to be less than
the expected signal although still not at high significance. One may be inclined
to conjecture that this could be due to the negative contribution coming from
the thermal SZ effect, but at this redshift $100\arcmin$ corresponds to $\approx 20
~h^{-1}$Mpc which would be too large a scale to be caused by hot gas in galaxy clusters.
Although the clusters do cluster among themselves, the contribution to any extend SZ 
effect is likely to be small \citep{Myers04}. Besides, there is no physical reason why 
should SZ effect only affect the highest redshift sample. The most likely explanation for 
this appears to be a statistical fluctuation which means that our SDSS LRG measurement 
rejects neither the ISW expectation nor the zero correlation at more than $\approx
1\sigma$ significance level. If we bin the data in the angular range
12\arcmin--120\arcmin ~into a single bin, the correlation amplitude and its
jackknife error ($1\sigma$) is $0.25\pm 0.33$ $\mu$K which deviates from the null
result hypothesis by only $0.8\sigma$ and from the standard model by 
$1.0\sigma$. For the 2SLAQ case, as in other studies, the positive
cross-correlation signal agrees very well with the expected ISW signal in the
standard cosmology in terms of its amplitude and angular extent. Nevertheless,
the 2SLAQ  sample's rejection of the null result is still only at the
1.2-1.6$\sigma$ significance level (see Table \ref{tab:significance}).

\subsection{The Combined LRG sample}
\label{sec:combined}

We shall now consider the cross--correlation of the combined LRG sample with the
CMB data. In our first method of combining the three LRG samples we shall treat
these as three independent surveys and then test this assumption by presenting
the cross-correlation result for the combined 1.5 million LRG sample, complete
with its own direct jackknife error analysis, to check that they agree.

First, the three CCF's of the SDSS, 2SLAQ and AA$\Omega$ samples are combined by
weighting inversely in quadrature according to the statistical errors of each
sample (see bottom right panels of Fig. \ref{fig:mod} and also Fig.
\ref{fig:combined}). The model (red solid line in Fig. \ref{fig:mod}) is
estimated by taking an average of the ISW models of the three LRG samples. We
find that the rejection significance is 1.3$\sigma$ for the standard ISW model
and 1.0$\sigma$ for the null result in the  $W$ band. In the ILC band the
significance of the rejection of the ISW model rises to 2.0$\sigma$ and the
significance of rejection of the null result reduces to 0.5$\sigma$.  Table
\ref{tab:significance} gives the summary of all the significance tests
performed. We conclude that while the ISW standard model is still consistent
with the CCF result from the three combined, weighted  LRG samples it is now
more consistent with the null result due to the inclusion of the AA$\Omega$
data.

Second, for comparison,  we also present the  results of cross-correlating the
combined LRG sample with the WMAP5 data i.e. we now treat the combined sample of
$\approx$ 1.5 million LRGs as a single sample for cross-correlating with, in
turn, the WMAP5 $W$ and $ILC$ CMB data. A full jackknife error analysis was
carried out in the same way as for the individual samples. We expect the results
to be similar to the weighted combination of the three samples' CCF's as
presented above. Fig. \ref{fig:combined} shows the comparison between these
results. The models for the combined samples  are computed following the
procedure described in \S \ref{sec:theory} asssuming the linear galaxy bias (given in
Table \ref{tab:significance}) estimated from the angular autocorrelation
function and $N(z)$ of 1.5 million LRGs.  Table \ref{tab:significance} again
shows the significances of rejection of the standard model and the null results.
We see that the observational  results in both cases are very similar.  For both
bands, the significances are given in Table \ref{tab:significance}. The results
are again very similar to those where the weighted mean was adopted. The
cross-correlation results are again as consistent  with the zero correlation as
they are with the standard ISW model  for the $W$  band. The ILC band again more
signifcantly rejects the ISW model than the null result.

\begin{figure*}
\hspace{-0.4cm}
	\centering
 \includegraphics[scale=0.5]{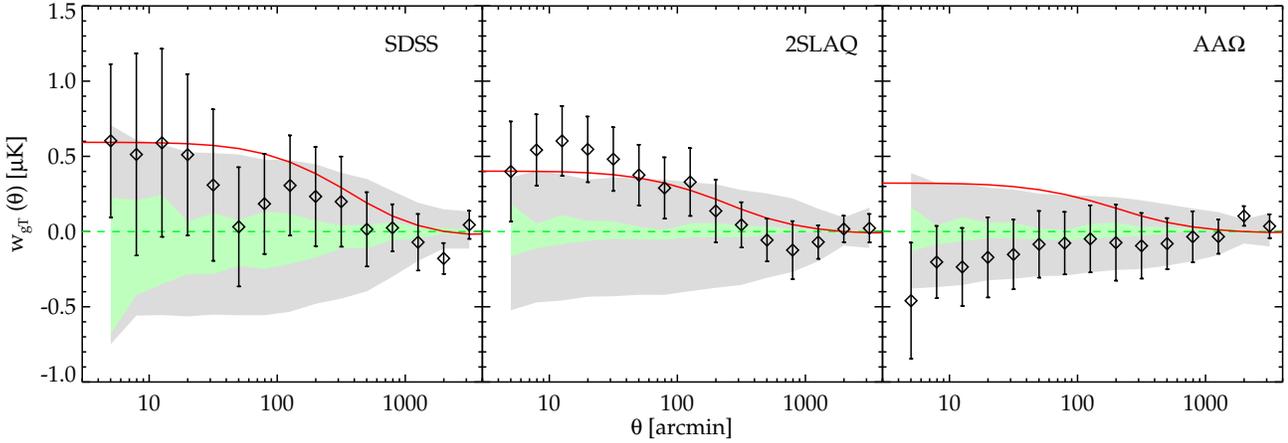}

	\caption{The cross--correlation results (diamonds) of three LRG samples and
	their jackknife errors ($1\sigma$) compared to the results of using 100
	random realisations of each LRG sample (inner green shaded region) and 100
	simulated CMB maps (outer grey shaded region). The shaded area signify a
	standard deviation in the measurement of 100 realisation for each case. Note
	that the means of these random realisations are consistent with zero as can
	be seen from their symmetry about the zero line. The solid line is again the
	theoretical prediction of the ISW signal in standard $\Lambda$CDM.}
	\label{fig:rand}
\end{figure*}

Clearly the preference for the null result over the standard model prediction
depends on the accuracy of the new AA$\Omega$ result. We test the robustness of the
AA$\Omega$ result in \S \ref{sec:robustness}.

\subsection{$\chi^2$ fits}
\label{sec:chi2}

For completeness, we also quantified the goodness-of-fit of our measurements to
the expected ISW signal or null result hypothesis by calculating the
\textit{chi-square}, $\chi^2$, which uses the normal size bin as shown in Fig.
\ref{fig:mod} and takes into account the possible correlation of the bins
through the use of the covariance matrix (\S \ref{sec:technique}). The $\chi^2$
is given by

\begin{equation}
 \chi^2=\sum_{i,j} \mathbfss{C}_{ij}^{-1}[w_{gT}(\theta_{i})-w_{gT}^{\rm{ISW}}(\theta_i)]\cdot[w_{gT}(\theta_{j})-w_{gT}^{\rm{ISW}}(\theta_j)]
\end{equation}
where $\mathbfss{C}_{ij}^{-1}$ is the inverse of covariance matrix,
$w_{gT}(\theta_{i})$ is the measured angular cross--correlation and
$w_{gT}^{\rm{ISW}}$ is the theoretical expectation assuming the standard
$\Lambda$CDM cosmology (see Fig. \ref{fig:mod}) which can be replaced by zero
when testing the zero correlation hypothesis. Using the galaxy linear bias, $b_g$, and
$N(z)$ for each sample as as mentioned in \S \ref{sec:comparison}, the $\chi^2$ tests were performed for
the angular bins between 12\arcmin--120\arcmin, inclusively. The lower limit is
set approximately to the best WMAP5 resolution in the \W band ($\approx 12\arcmin$).

The significances obtained from the $\chi^2$ method generally  confirmed the
results using the 12\arcmin--120\arcmin~bin, especially those of the main LRG
samples. For example, assuming  standard model parameters, the SDSS-W band
results give $\chi^2$=19.4 for the predicted ISW signal and 17.7 for the zero
correlation hypothesis. For the 2SLAQ results, using the standard model gives
$\chi^2=13.2$ and relative to the null result gives $\chi^2=11.5$. These
$\chi^2$ were computed for 6 degrees-of-freedom (d.o.f). Using the $\chi^2$
distribution, the SDSS results deviate from the ISW model and null result at 4
and 7 per\,cent statistical significance, respectively. The 2SLAQ results agree
with the ISW model with the reduced chi-square, $\chi^2$/d.o.f., of order of
unity and reject the zero correlation hypothesis at $1.5\sigma$ significance.
The AA$\Omega$ results gave  $\chi^2=11.7$ and $\chi^2=4.4$ for the ISW model
and null correlation respectively, corresponding to the chances of agreement of
7\% and 62\%. These all  agree reasonably well with the large-bin significances
presented in Table \ref{tab:significance}. However, the similar $\chi^2$
significance tests of the combined sample and some ILC individual samples did
not perform very consistently, occasionally giving pathological results and poor
agreement with the 12\arcmin--120\arcmin~bin and this is why we have only  
quoted the simpler, single large-bin significances in Table
\ref{tab:significance}.

\section{Robustness tests}
\label{sec:robustness}
Given that the AA$\Omega$ LRGs comprise a new sample, there is no previous measurement that
can be directly compared to our own. We now present the result of tests we
have done in order to check the robustness of our new result.

\subsection{Random realisations and simulated CMB Maps} 
Here we generate 100 random realisations for each of the sample. Each realisation 
has the same number density as the sample it tries to mimic. Note that these random 
realisations are unclustered. The results are shown in Fig. \ref{fig:rand}. 
The jack-knife errors that we use are seen to be much larger than the standard deviation 
of the random catalogues (inner green shaded region). This is expected because the random 
catalogues are unclustered unlike the LRGs. The means of these random realisations are consistent
with zero and show no sign of bias except perhaps at the smallest scales of the SDSS sample.

We have also made simulated CMB temperature anisotropy maps and cross-correlated
these with the three LRG samples. A simulated CMB map is generated as a
realisation of random gaussian fields on a sphere with the fluctuation
characterised by \textit{WMAP}5 best-fit power spectrum. The simulated maps are
also convolved with a Gaussian beam with FWHM similar to the \textit{WMAP}
$W$-band, i.e. 12\arcmin.6. The cross--correlation results are shown in Fig.
\ref{fig:rand}. The standard deviation of 100 CMB random realisations (outer
grey shaded region) are roughly consistent with our jackknife estimates
especially at small and intermediate scales but somewhat larger at large scales.

\begin{figure}
\hspace{-0.6cm}
\centering
\includegraphics[scale=0.5]{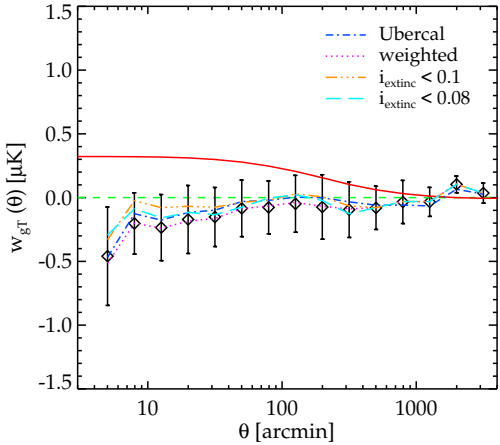}

	\caption{The cross-correlation of the AA$\Omega$ LRG to W-band data using the original SDSS photometry (diamond) compared to the measurements using `ubercalibration' (dot-dash line), the stripe weighted (dotted line) and when the data is restricted to the region where galactic extinction in $i$-band less than 0.1 (dot-dot-dot-dash line) and 0.08 (long dashed line) magnitude.}
	\label{fig:ao_weight}
\end{figure}

\subsection{Photometry test}
\label{sec:photo}
Next, we look to see if the AA$\Omega$ cross-correlation measurement is robust by comparing the result from the
SDSS `ubercalibration' of \citet{Padmanabhan08} with that from the standard SDSS
calibration. Fig. \ref{fig:ao_weight} shows that the results are stable to
whichever calibration we used. We further looked for systematic effects in the
original photometry by weighting SDSS stripes according to their overall number
density. The physical motivation for this  arises from  the SDSS observing
strategy and the fact that a slightly different calibration for different nights could
affect the source density as a function of the SDSS stripe, given our faint
limit. We observe a hint of these variations although not at a high level and
use these to correct the source densities in each stripe as mentioned. However,
such variations seems to be weaker when using the ubercalibration as opposed to
the standard one. The result of weighting according to the stripe number
density is shown in Fig. \ref{fig:ao_weight} and again the result appears
robust when this filter is applied to the original data.

Although we work at a relatively high galactic latitude, it is possible that in 
some regions of the sky, high  galactic dust obscuration could result
in lower detections of faint objects. Furthermore, that same dust obscuration patch could 
be a source of contamination in the CMB data in the sense that the temperature in that
particular region could be systematically raised by the dust emission and hence results
in a false anticorrelation. To test this, we exclude the region where the extinction
is greater than 0.1 mag in the $i$-band which discards $\approx15$ per\,cent of the data. 
We observe no difference to our main results, even when a more aggressive limit, 
$i_{\rm extinction} < 0.08$ (23 per\,cent discard), is applied 
(see Fig. \ref{fig:ao_weight}). Note that when similar tests are performed using 
extinction in the SDSS $r$-band instead, we again obtain results which are consistent with 
those presented in \S \ref{sec:result} for all three LRG samples.   

\begin{figure}
\hspace{-0.6cm}
	\centering
\includegraphics[scale=0.5]{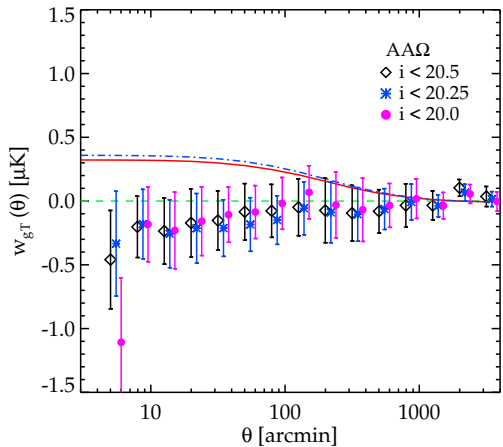}

	\caption{The cross-correlation of the AA$\Omega$ LRG to W-band data compared to the measurements using the same colour--colour selection sample but with brighter faint--limit cut, $i_{\rm{deV}}<20.25$ and $i_{\rm{deV}}<20.0$. Only the theoretical expectation of the full (solid line) and $i_{\rm{deV}}<20.25$ (dash-dot line) sample are shown. The data points are shifted slightly for displaying purposes.}
	\label{fig:ao_mag}
\end{figure}

We then cut back the $i$-band limit of the AA$\Omega$ sample in 0.25 mag steps
from $i=20.5$ to $i=20.0$ while keeping the other conditions the same. These
results are compared with the result at $i<20.5$ in Fig. \ref{fig:ao_mag}. Again
the results appear robust. We have also made tests of the single epoch SDSS
photometry using deeper Stripe 82 \citep{Abazajian09} and the William Herschel Deep
Field (WHDF, \citealt{NM01}) data. Both these comparisons showed that the SDSS
photometry in $r$, $i$ and $z$ bands showed good agreement with the deeper
data until the errors showed a significant increase beyond the limits $r=22.0$, 
$i=21.0$ and $z=20.2$.

\begin{figure}
\hspace{-0.6cm}
\centering
\includegraphics[scale=0.50]{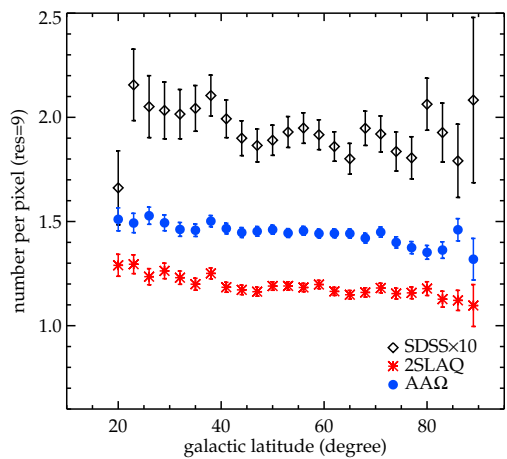}

	\caption{The object numbers per pixel as a function of galactic latitude, $b$. Recall that we use equal area ($\approx 49 ~\rm{arcmin}^2$) pixels with res=9 resolution scheme \citep[HEALPix,][]{HEALPixref}. The SDSS number has been multiplied by 10 to extend the plot range.}
	\label{fig:numb}
\end{figure}

\begin{figure}
\hspace{-0.6cm}
\centering
\includegraphics[scale=0.45]{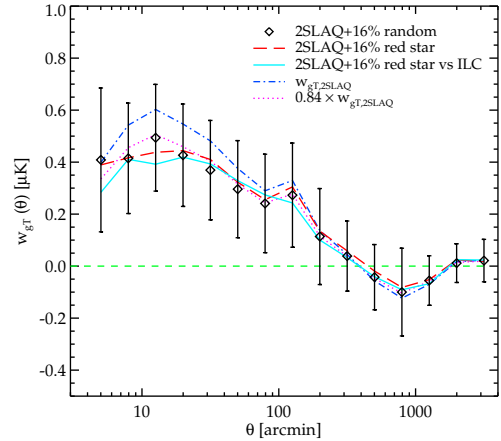}

	\caption{The cross--correlation of the $W$-band data and the 2SLAQ LRG when a sample of random realisation of $\approx$16 per\,cent is added to the LRG catalogue (diamonds). The results using original 2SLAQ sample and when multiplied by $1-f_{s}$ are shown as the dot-dashed and dotted lines, respectively. The long-dashed line shows the result when the 16 per\,cent added contaminants are replcaed by red stars. The result of cross--correlating the ILC map with the 16 per\,cent--red star contaminated 2SLAQ sample is also shown (solid line).}
	\label{fig:slaqaddrand}
\end{figure}

\subsection{Star--galaxy separation}
\label{sec:stellar}
We  noted in \S \ref{sec:data} that the stellar contamination in our
AA$\Omega$-LRG sample could be as high as 16 per\,cent. Care should therefore be
taken when analysing this dataset. We obtained this contamination fraction using
the information learned from the AA$\Omega$-LRG spectroscopic survey
\citep[Sawangwit et al. in prep]{Ross08}, by imposing a star--galaxy separation
in the  $z$-band similar to the method applied in the SDSS- and
2SLAQ-LRG redshift surveys using  the $i$-band. Our optimised
star--galaxy separation procedure selects objects with $z_{\rm psf}-z_{\rm
model} > 0.53+0.53(19.0-z_{\rm model})$ which only loses  genuine LRGs at a
sub-percent level and leaves $\approx$16 per\,cent stellar contamination in our
sample, as quoted eariler.

The effect of stellar contamination distributed at random 
in the sample is simply a dilution of the
over/under density hence reducing the autocorrelation amplitude of the sample by
$(1-f)^2$ and the cross--correlation by a factor of $(1-f)$ where $f$ is the
fraction of the contamination. This is particularly true if the contamination is
distributed  uniformly at random in the sample. However, if there is some
spatially dependent variation of the number density, a further systematic effect
could arise through this process. To test this, we first check to see if there
is a trend of the number density as a function of galactic latitude as one might
expect for stellar contamination. Although a slight such trend is observed, it
is at no more than the levels observed in the SDSS and 2SLAQ samples (see Fig.
\ref{fig:numb}) whose stellar contamination fractions are approximately 1 and 
5 per\,cent, respectively. Next we restrict the data to the high galactic 
latitude regions, namely $b > 40\deg, 50\deg$ and $60\deg$. The results are in
good agreement with our main results for all three LRG samples up to $b > 60\deg$
where the cross--correlations become noisy due to the 75 per\,cent reduction in 
the sample sizes. 

\begin{figure*}
	\centering
 \includegraphics[width=85mm]{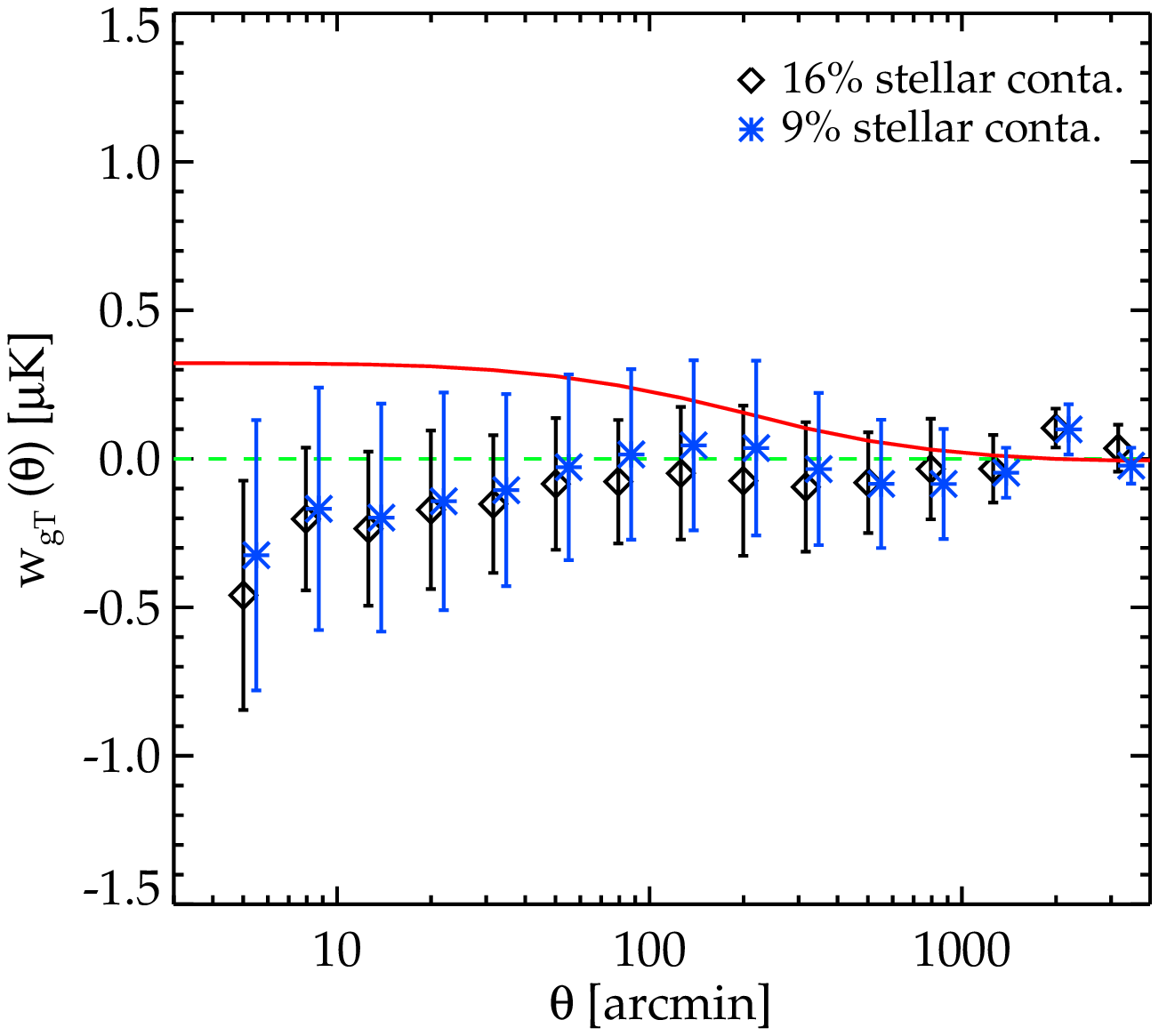}\includegraphics[width=85mm]{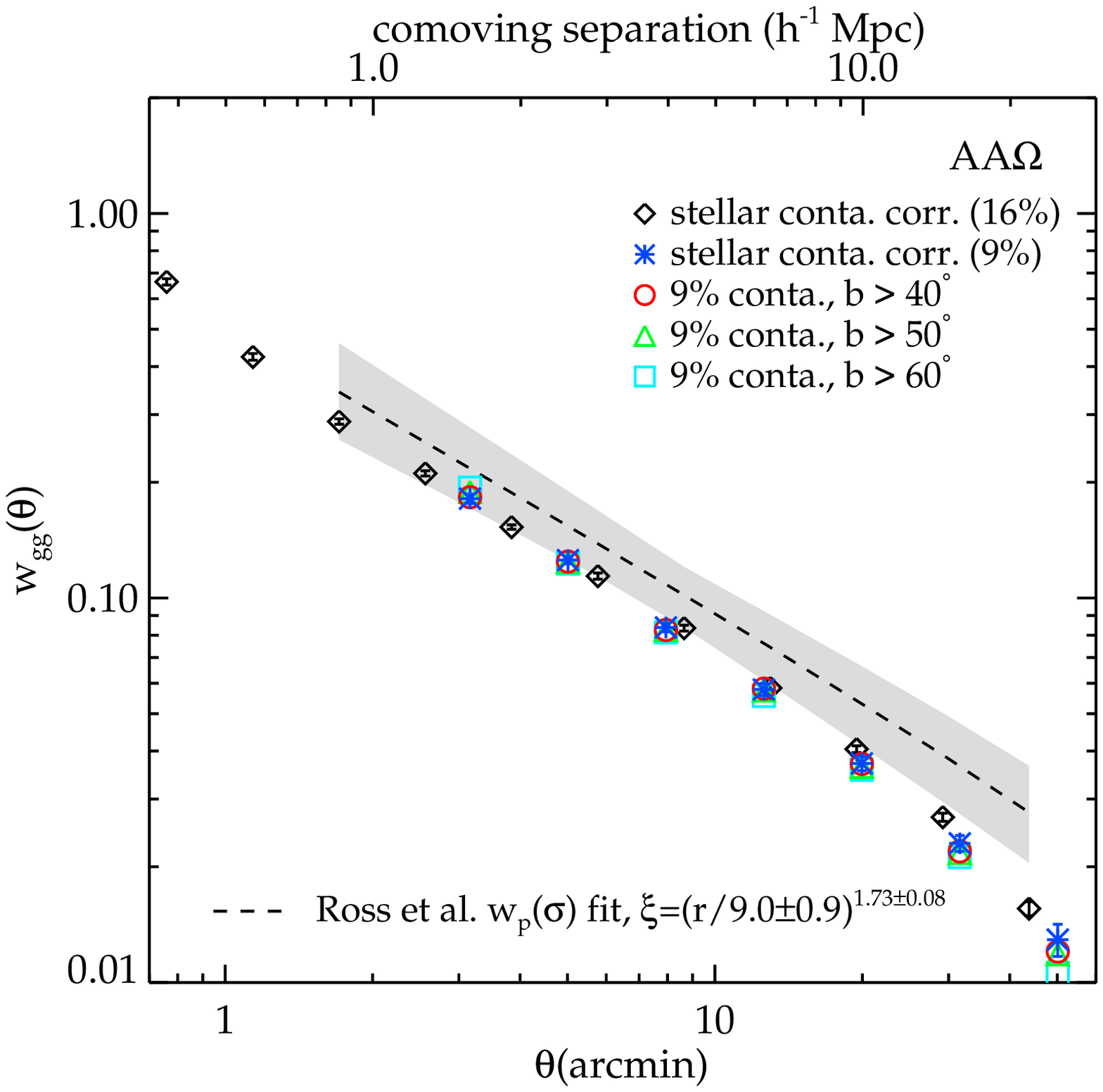}
\caption{Left: The AA$\Omega$ LRG--WMAP5 cross--correlation of the 9\% stellar contaminated
sample (asterisks) compared to the main AA$\Omega$ sample used in our study (diamonds). Right: The 
corrected autocorrelation functions of the 9\%-- and 16\%--contaminated samples (asterisks and diamonds).
These are compared to the results of limiting the 9\%--contaminated sample to the regions with galactic 
latitude higher than $>40^\circ$, $50^\circ$ and $60^\circ$. The dashed-line and shaded region is the 
$w_{gg}(\theta)$ and $1\sigma$ error inferred from the $w_{p}(\sigma)$ 
measured from $\approx 400$ spectroscopically confirmed AA$\Omega$ LRGs \citep{Ross08}.}

\label{fig:9percent}
\end{figure*}

To simulate the effect of the stellar contamination on the LRG--CMB cross--correlation, 
we have introduced a set of random realisations into the 2SLAQ sample. The result
is presented (diamonds) in Fig. \ref{fig:slaqaddrand} along with the cross--correlation of
the original 2SLAQ sample (dot-dashed line) and the result of reducing its amplitude by a factor
of $(1-0.16)$ (dotted line). Furthermore we would like to check for any effects that may arise from
possible large scale clustering of the stars. This is done by adding a sample of red stars to 
the 2SLAQ LRG sample at the 16 per\,cent level. The stars are selected 
with similar colour--magnitude criteria to that of the AA$\Omega$ LRGs and should mimic 
the angular distribution and properties of the stellar contaminants seen in the sample.
The result is shown in Fig. \ref{fig:slaqaddrand} (long-dashed line). This test should also reveal
any possible effects on the $w_{gT}$'s due to (if any) correlation between these stars and the CMB.
We found the 16 per\,cent--red star contaminated 2SLAQ result to be consistent with the dilution of
randomly distributed contaminants case. The result is also consistent with the cross--correlation with
the foreground reduced ILC map (solid line), further confirming that our result is not affected by 
any star--CMB cross--correlation. Note that the significance test presented in Table
\ref{tab:significance} has already taken into account such an effect by
multiplying the ISW model by a factor of $(1-0.16)$. The significance of the
AA$\Omega$ sample's rejection of the standard model ISW prediction is therefore
robust against the stellar contamination discussed here.

We next attempt to reduce the stellar contamination fraction by imposing 
a more aggressive star--galaxy separation cut which result in
nearly halving the number of genuine AA$\Omega$ LRGs. The cut is a combination of the fitted `de Vaucouleurs' 
radius as a function of $z_{\rm deV}$ magnitude and the correlation between the `de Vaucouleurs' and fiber 
magnitudes in $z$-band. This reduces the contamination to $\approx$9 per\,cent. Fig. \ref{fig:9percent} (left panel) 
shows the cross--correlation of this sample with the $W$-band data which is in good agreement with our main result.

The contamination fractions of these samples are verified by their angular autocorrelation functions, 
$w_{gg}(\theta)$. The corrected $w_{gg}(\theta)$ is shown in right panel of Fig. \ref{fig:9percent}. 
This is again in good agreement with the 16 per\,cent contaminated sample and consistent within $\approx1\sigma$
of the \cite{Ross08} power--law fit to the semi--projected correlation function, $w_{p}(\sigma)$. 
Note that we only expect the agreement in the range $r\approx$1--15$\hmpc$ where a single power-law is a good fit 
to the data. The measured $w_{gg}$'s are also consistent with the results when restricting galactic latitude to greater 
than $40\deg$, $50\deg$ and $60\deg$. We believe the slight discrepancy with the $w_{p}(\sigma)$ is due to the 
noisy measurement from the small number of spectroscopically confirmed LRGs used in \cite{Ross08} and not 
caused by the under--estimation of the contamination level as demonstrated by our two independent 
approaches for star--galaxy separation.

Even if the contamination fraction is under-estimated, the effect of an increased
(uniform) stellar contamination would be to increase the ISW model amplitude
when the bias value from the LRG autocorrelation is corrected upwards to obtain
the true bias value. This upwards shift in the ISW model would then be exactly
cancelled by the downwards correction to account for the dilution of
the cross-correlation signal due to stellar contamination.

We conclude that despite the faint magnitude limit, and moderate level of
stellar contamination ($\approx16$ per\,cent) our ISW results for the AA$\Omega$
LRGs seem robust to the tests we have made and the SDSS data seem accurate
enough to support this ISW analysis. Up to this point, we have therefore found
no explanation in terms of a systematic effect for the low
AA$\Omega$--\textit{WMAP}5 cross--correlation result. Next, we shall perform a
similar analysis on some of the large-scale tracers whose ISW effect has been
previously claimed in order to test our methodology and look for other possible
systematics in these samples.

\begin{figure*}
	\centering
 \includegraphics[scale=0.5]{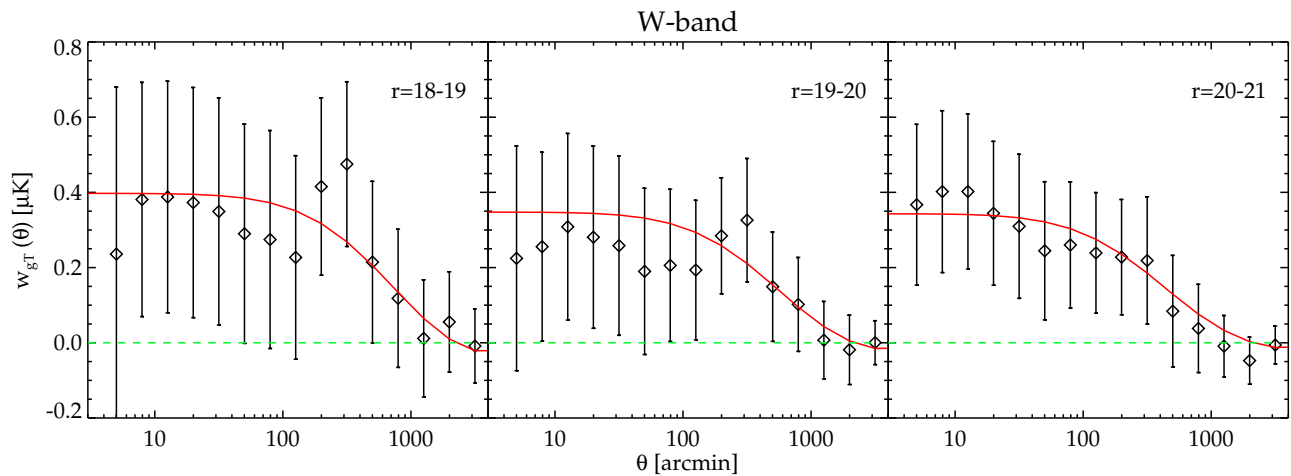}
 \caption{The cross--correlation of \textit{W}-band data with the $r$-band
	selected galaxies. The sample magnitude ranges are as indicated in the plots.
	The ISW model prediction is shown for each sample, assuming
	\citet{Dodelson02} $n(z)$ with the measured bias of 1.2, 1.1 and 1.2 for $
	18 < r < 19$, $19 < r < 20$ and $20 < r < 21$ sample, respectively.}
	\label{fig:rband}
\end{figure*}

\subsection{SDSS galaxy--\textit{WMAP}5}

We next cross--correlate galaxies extracted from SDSS DR5 using  $r$-band
magnitude limits. The objects are photometrically classified as galaxies by the
SDSS reduction pipeline. We subsample the galaxies in three magnitude ranges,
namely, $18<r<19$, $19<r<20$ and $20<r<21$, where all the magnitudes are
galactic extinction corrected model magnitudes. The subsamples contain
approximately 2, 6 and 16 millions objects, respectively. This is the same as
\citet[C06 hereafter]{Cabre06} but covering $\approx$ 20 per\,cent more area and
we use \textit{WMAP}5 rather than \textit{WMAP}3. A similar $r$-band selected
galaxy sample was also used by \citet[G08 hereafter]{Giannantonio08} although
these authors use `ubercalibration' photometry rather than the original one and
limit the sample photo-z to redshift between 0.1--0.9. The ISW effect has been
claimed to be detected in these samples at moderate significance level by both
C06 and G08 although their results do not agree with the former having twice as
much positive cross--correlation between the CMB and the $r$-band selected
galaxy sample. As a result C06 need to fit their result with higher
$\Omega_\Lambda$, for a galaxy bias $b=1.0$.

For the cross-correlation analysis, we proceed in the same manner as with the LRG samples. 
To compute the ISW model, we use the $n(z)$ distributions following \citet{Dodelson02}.
The average redshifts inferred from the $n(z)$ are estimated to be 
approximately 0.17, 0.24 and 0.33. We then follow our procedure for the LRGs 
and obtain the galaxy linear bias from the measured amplitude of
the galaxy 2-point autocorrelation function for each subsample. We obtain the values 
$b_g=1.2, 1.1$ and 1.2 for the sample with $r$-band magnitude limit of 18--19, 19--20 and 20--21, 
respectively, in agreement with the measurements of C06 and G08 whose
$b_g \approx$1--1.2. The cross--correlation measurements and the theoretical models are 
presented in Fig. \ref{fig:rband}.

We marginally detected the correlation between the CMB data and all the $r$-band
selected subsamples. We shall now compare the $20 < r <21$ result in  Fig.
\ref{fig:rband} to Fig. 2 (top) of C06. Our result is lower by a factor of
$\approx2$ but very close to the re-analysis of the SDSS $r$-band data of G08
who also found a factor of two discrepancy with C06. After their discussions,
the two groups found that the discrepancy is due to an extra quality--cut
imposed on the data by C06, namely, $r$-band magnitude error less than 0.2 mag.
We regard the factor of two rise in the amplitude of the cross-correlation after
this small change in the magnitude error limit simply as symptomatic of the
statistical fragility of the result. We conclude that our re-analyses of these data agree well with
the standard $\Lambda$CDM predictions although the significance of rejection of
the null result is still only $\approx$1--2$\sigma$.


\subsection{NVSS-\textit{WMAP}5 cross--correlation}

To test our methodology further, we performed a cross--correlation analysis of
\textit{WMAP}5 with radio sources from the NRAO VLA Sky Survey (NVSS;
\citealt{Condon98}) which has been previously  used by various groups for ISW
studies. The NVSS sample comprises about 1.8 million radio sources detected to a
flux limit of $\approx$ 2.5mJy at 1.4GHz. The NVSS covers the entire sky higher
than $-40\deg$ declination ($\approx$80 per\,cent of the sky). Interestingly, the
previous study of \citet{Boughn02} found no correlation of these sources with
the Cosmic Background Explorer (COBE) CMB map but a later study by \citet{Nolta04}
found a positive correlation with the first-year \textit{WMAP} data which they
claimed to be the evidence for $\Omega_\Lambda > 0$ at 95\% confidence, assuming a flat
CDM cosmology. The re-analysis of the NVSS--CMB correlation by G08 also
confirmed \citet{Nolta04} results at approximately the $3\sigma$ significance level.
 
For the cross--correlation analysis we restrict the data to the declination, 
$\delta \ge -37\deg$ where the survey is the most complete. We then applied the 
masking and pixelisation procedure described in \S \ref{sec:technique} but
for this sample we shall use lower resolution (res=6 as opposed to res=9)
HEALPix \citet{HEALPixref} scheme to reduce the computing time because of
the much larger sky coverage of the NVSS. We checked that the measurements using
different resolutions do give the same results in terms of amplitudes and statistical 
uncertainties (\S \ref{sec:robustness}). The higher resolution (res=9) result 
shall be discussed in this section but for the purpose of the systematics test in \S
\ref{sec:rotate}, we shall present the results using res=6. 

\begin{figure}
\hspace{-0.5cm}
	\centering
\includegraphics[scale=0.5]{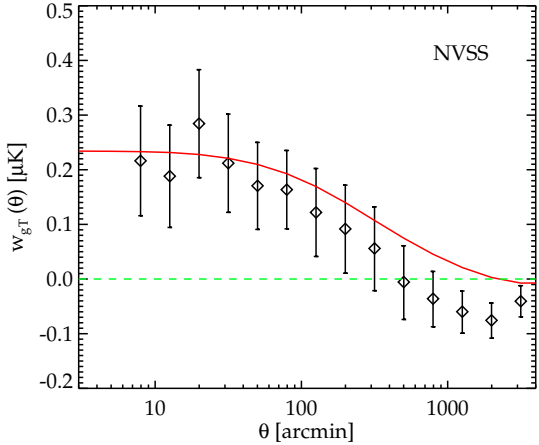}

\caption{The cross-correlation of the NVSS sources to the \textit{W}-band data. The ISW prediction (red solid line), assumes linear bias of 1.5 \citep{Boughn02,Giannantonio08} and $n(z)$ derived from \citet{Dunlop90} radio source luminosity function (mean-$z$ model 1).}
	\label{fig:nvss}
\end{figure}

\citet{Boughn02} noticed a number density trend with the
declination which affected their autocorrelation measurement. 
Following \citet{Nolta04}, we applied a  correction for this by splitting the sample into
$\sin{\delta}$ strips of width $\approx 0.1$ and scaling the galaxy numbers in pixels
belonging  to a particular strip by the ratio of global mean to the strip mean. 
The cross--correlation procedure is then carried out as outlined in \S \ref{sec:technique}
but the statistical uncertainties and covariance matrix are now estimated from 
approximately 20 equal--area  jackknifes rather than 24. The result using 
\textit{W}-band data along with the standard model ISW prediction (red solid line) 
is presented in Fig. \ref{fig:nvss}. 

The ISW predictions for the NVSS sources are computed using the number--redshift
distribution, $n(z)$, derived from the radio source luminosity function (mean-$z$ model 1)
of \citet{Dunlop90}. The median redshift estimated from such $n(z)$ is $\approx0.8$
with a tail extending out to $z \approx 3$. We assume the source bias, $b$, of 1.5 as measured 
by a number of authors \citep[e.g.][]{Boughn02,Giannantonio08}.

Fig. \ref{fig:nvss} shows that we find a marginally positive correlation similar to
the prediction of the standard model at scales smaller than $\approx 5\deg$ at $\approx 2\sigma$
significance. Our result can be directly compared with that of G08 who, like us, cross--correlate the 
source number fluctuations to $\Delta_{\rm{T}}$ as opposed to the source number per pixel
approach of \citet{Nolta04}. Similary, G08 observed a good agreement between their measurement and 
the standard $\Lambda$CDM model which also starts to breakdown at $\approx 5\deg$.
We take this agreement as a further indication of the robustness of our cross--correlation 
methodology and analyses. In the next section, we shall further test the NVSS--\textit{WMAP}5 result for contamination
by systematic effects. 

\begin{figure}
\hspace{-0.5cm}
	
\includegraphics[scale=0.48]{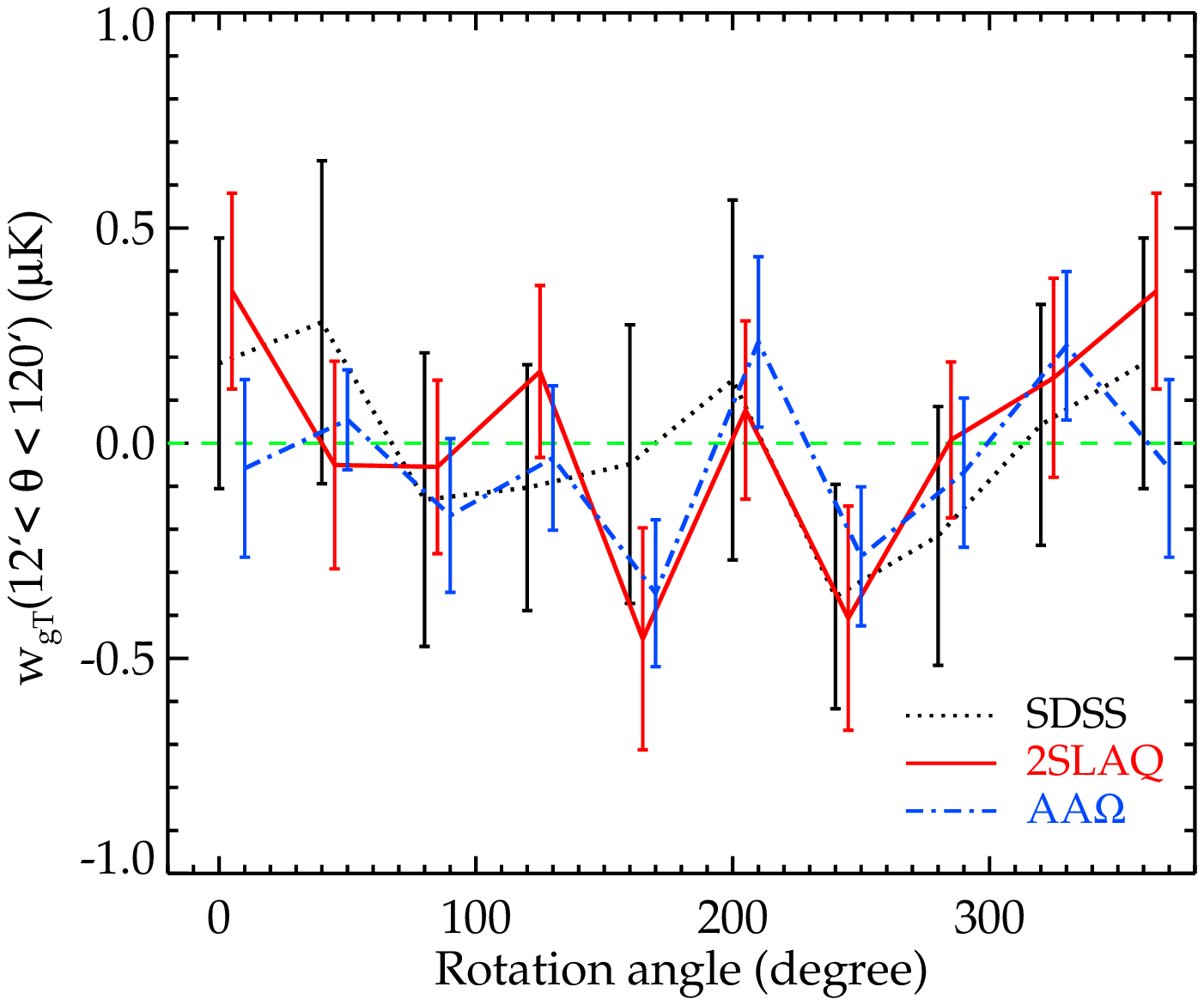}
\hspace{-0.5cm}

\includegraphics[scale=0.48]{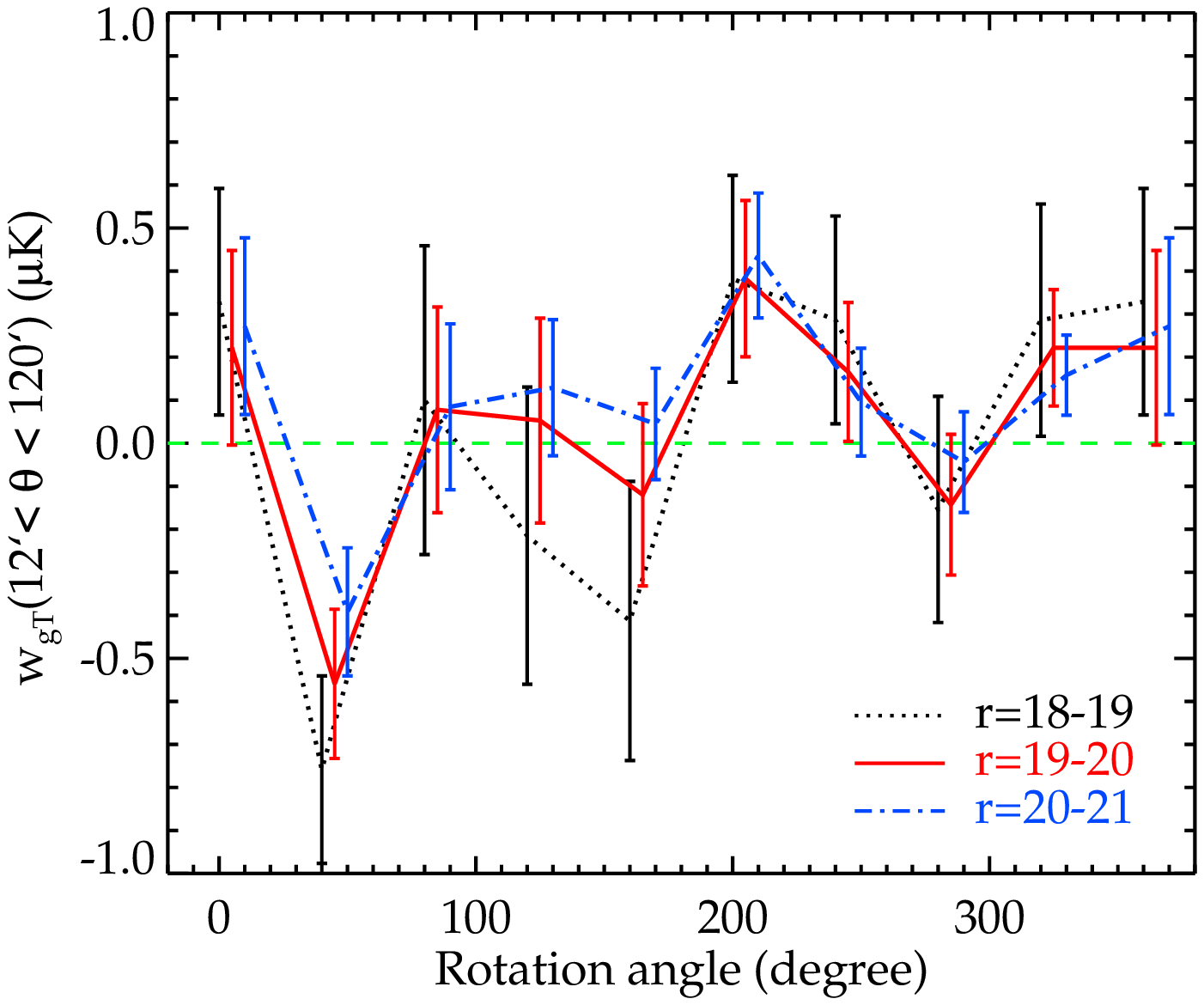} 
\hspace{-0.5cm}

 \includegraphics[scale=0.48]{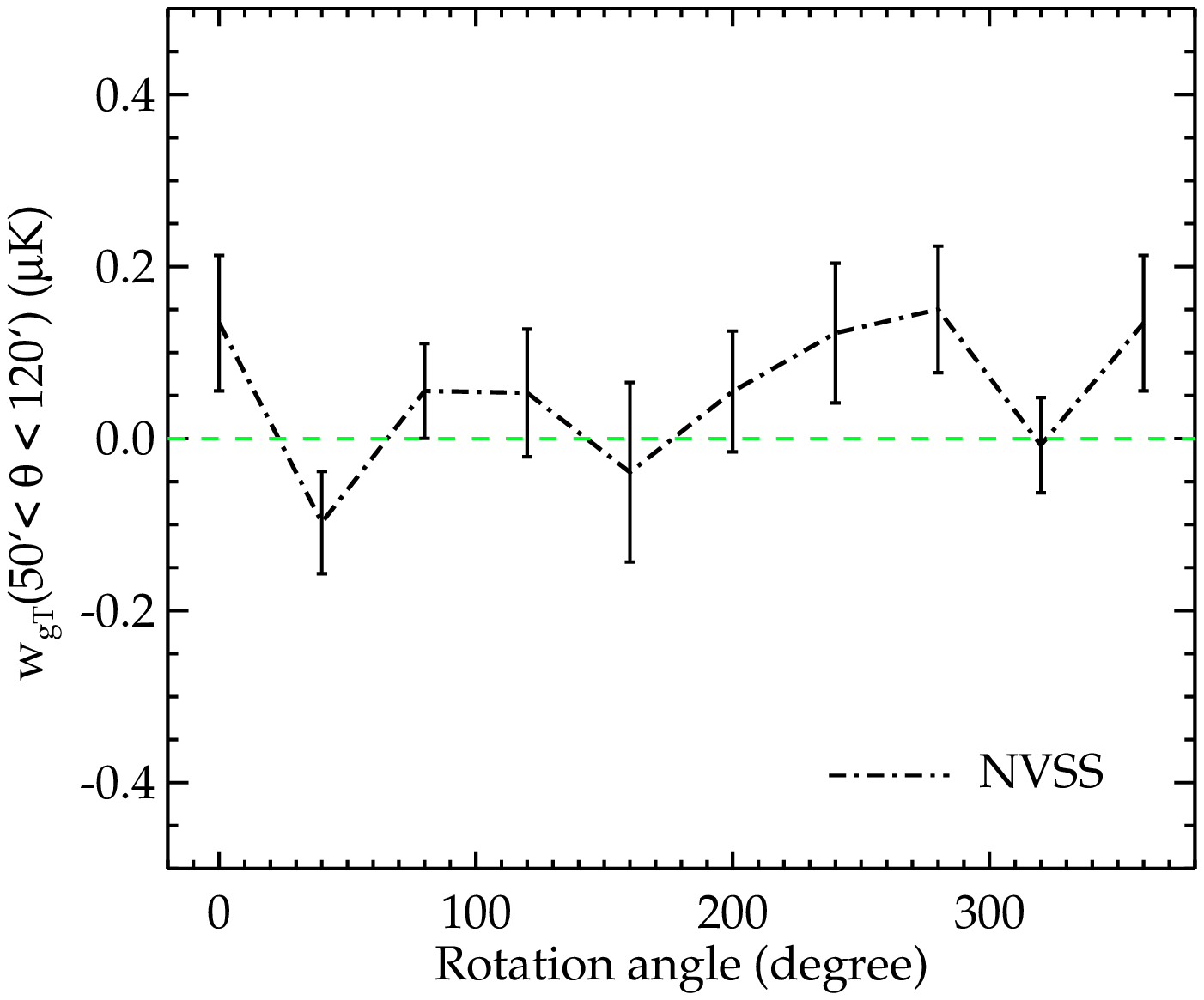}

	\caption{The cross-correlation of the three LRG samples (top), $r$-band selected galaxies (middle) and the NVSS sources (botom) to the rotated \textit{W}-band data in our rotation test (see text for more details). Note that for the top two plots, the points have been shifted slightly in the x-axis for clarity.}
	\label{fig:rotate}
\end{figure}

\section{CMB Sky rotation test}
\label{sec:rotate}
Here we shall perform an additional test for systematics, similar to
that used by \cite{Myers04} and \cite{Bielby07} for testing their detection 
of the SZ effect, particularly in checking the reality of a large
scale temperature decrement around galaxy groups and clusters. 
We follow these authors and rotate the \textit{WMAP} maps
around the galactic pole in the clockwise direction, each time adding $40\deg$
to galactic $l$. There is an area very close to the pole where
there is less movement from the rotation, but given that we use a $40\deg$
shift the effect of this slight non-independence is small. We have
checked that if we cut out the circum-polar region down to galactic
latitude $b=75\deg$ our results are unaffected.

The CMB masks (KQ75 plus point source) are rotated with the temperature 
maps to ensure that the contaminated regions are excluded from 
both galaxy and temperature fluctuation maps. The SDSS DR5 mask is then 
applied to the data in the case of LRG and $r$-band selected samples. 
The cross--correlation is performed using the \textit{W}-band data following
the procedure described in \S \ref{sec:technique}. 
We use the cross--correlation results between $12\arcmin < \theta < 120\arcmin$ 
where the difference between the ISW and null result is at its maximum as  
in \S \ref{sec:result}. The cross--correlations are then performed at 
eight $40\deg$ intervals. 

\subsection{LRGs}
The cross-correlation measurements are presented in Fig. \ref{fig:rotate} (top
panel). The errors shown are jackknife errors ($1\sigma$) and as expected they
are similar at all rotation angles which makes the data points straightforward
to compare. For the SDSS sample at z=0.35, there is a higher positive point at
rotation angle 40 degrees. For the 2SLAQ sample at z=0.55 the points at rotation
angles 160 and 240 degrees are more negative than the zero degree point is
positive. There is no reason to expect anything other than a null result at any
rotation angle other than zero. Therefore, based on this  rotation test the
significances are now reduced to the $\approx12-25$\% level, suggesting that
systematics as well as statistical errors may be affecting the data.

\subsection{SDSS galaxies}
We also applied the same test to the ISW results using three SDSS $r$-band selected
galaxy samples of $18 < r < 19$, $19 < r <20$ and $20 < r <21$. The results are
shown in Fig. \ref{fig:rotate} (middle). Again we see
that there are rotation angles that show more significant non-zero
detections than at the zero degree rotation angle. We see that at
$40\deg$ rotation angle, the results are very negative in all three samples.
At the rotation angle of $200\deg$, the results are more
positive than the zero degree rotation, again in all three samples. 
As for the LRG samples, this means that the significance is reduced to 
a marginal $\approx10$\% level and the results suggest that systematic 
effects as well as statistical errors may be contributing to the apparent 
ISW detection at zero degrees rotation angle.

\subsection{NVSS radio sources}
We then applied the same test to the NVSS--\textit{WMAP}5 cross--correlation result.
(see Fig. \ref{fig:rotate}, bottom). This time the point at rotation angle 280 degrees is
more positive than the point at zero degrees. As with other samples, the jackknife
errors on all the points are similar so this comparison is fair. Again
we conclude that systematic effects may be contributing to the apparent ISW detection 
which explains the reduction in statistical significance to $>10$\% from the rotation test.

\section{Discussion}
\label{sec:discuss}
Given the consistency of the AA$\Omega$ and the combined LRG results with the
zero correlation, we now discuss whether there is any contradiction between our conclusions 
and those of other authors. In particular, we discuss the results of G08 who claim 
a 4.5$\sigma$ ISW detection from the combined analyses of several large-scale tracers.
These tracers include some of the LRG samples. They also include NVSS radio sources. 
The most significant detection in their Table 1 is from the NVSS at 3.3$\sigma$. 
Their LRG analysis gives 2.2$\sigma$ for a sample roughly equivalent to our 2SLAQ 
LRG sample. These compare to 1.6$\sigma$ for our 2SLAQ samples. For the NVSS we find 
a $1.8\sigma$ result. Their SDSS galaxy sample gives $2.2\sigma$ equivalent to our combined
SDSS $r$--band limited sample which gives $\approx1.3\sigma$ significance. Thus our 
significances appear lower than those of G08, particularly for NVSS. 
This discrepancy increases when we consider the rotation test. In the rotation test of the NVSS sample,
1 out of 8 points has higher amplitude than zero rotation measurement which is only $\approx 1.5\sigma$ 
significance. For the 2SLAQ case, this gives 1--2 out of 8 points which is equivalent to 1.2--1.5$\sigma$ significance.
The SDSS galaxy gives 2 higher (or lower) points in 8 or $\approx 1.2\sigma$. 

However, G08 also have differences between their two methods of assessing the significance of their 
results. Their Table 1 assumes the hypothesis of the standard $\Lambda$CDM model to obtain a maximum
likelihood amplitude, $A$, and an associated error from their data. This error is different from the error
that can be inferred from the $\chi^2$ statistic in their Table 2 which tests the null result hypothesis. 
For example, their LRG result is 0.4$\sigma$ significant from Table 2 whereas it is 2.2$\sigma$ significant 
from Table 1. Their SDSS galaxy sample rejects the null result hypothesis at 1.3$\sigma$ significance from the 
$\chi^2$ statistic, again lower than their Table 1 at 2.2$\sigma$. Also the NVSS only reject the null 
result at 1.3$\sigma$ rather than 3.3$\sigma$.
We assume that these differences may be due partly to different null hypotheses and partly due to different 
methodologies. Certainly, the levels of significance in their Table 2 are lower and more in line with  
what our rotation tests show, i.e. 1--2 higher (or lower) points in 8 or 1.2--$1.5\sigma$. It remains to be seen 
for the other samples in their Table 1 and 2, if the same pattern applies with the maximum likelihood significances 
in Table 1 being higher than the $\chi^2$ significances in Table 2. We conclude that the rejection of the null 
result by their $\chi^2$ test may be more consistent with what we have found than the results in their Table 1. 
Indeed their $\chi^2$ summed from all surveys is 67 on 74 degrees of freedom which is hardly a significant rejection 
of the null result and can be compared to our overall rejection of the null result in our Table 2 of 0.5 to 1$\sigma$. 
Therefore as long as we refer to the $\chi^2$ test of G08, there seems to be no inconsistency with our estimate of 
the significance of the low rejection of the null result.

\section{Summary and Conclusion}
\label{sec:conclusion}
We have performed a cross--correlation analysis between the \textit{WMAP}5 CMB data
and various large-scale structure tracers including our new high redshift AA$\Omega$-LRG survey. 
The summarised conclusions of our findings are as follows:

\begin{itemize}

\item We have found a null ISW result for $z \approx 0.7$ AA$\Omega$--LRG sample.
The standard model is rejected at $\approx$ 3\% significance by this dataset.

\item We have confirmed the marginal correlations  between \textit{WMAP}5 CMB temperature
fluctuations and SDSS LRGs at z=0.35 and  2SLAQ LRGs at z=0.55. 

\item The null result in the AA$\Omega$--LRG sample at large scales
is unlikely to be caused by the negative contribution of the SZ effect, 
given its angular extent and the expected amplitude of ISW signal.

\item We have made a range of tests on the AA$\Omega$ cross-correlation measurement
which confirms its robustness. These include 
moving the magnitude limits up to 0.5mag brighter, removing areas of sky
with significant dust absorption, using an estimate of the cross-correlation
that takes out any possible systematic effects due to SDSS stripes and comparing
the standard and uber-calibrations of the SDSS photometry. We have also checked the 
effects of stellar contamination in our samples. All these tests produce results 
consistent with our original measurements.
 
\item We have also reproduced the cross-correlation results of most previous
authors using our techniques. In particular we have reproduced the marginally
positive correlations seen using SDSS magnitude limited samples of galaxies and
NVSS radio sources.

\item However, rotation tests indicate that accidental alignment or some unknown
systematics can give rise to a correlation signal comparable to and in many
cases even larger than the ISW signal itself. This suggests that the previous
positive detections may still be subject to unknown systematic effects.
 
\item Combining the new $\bar{z}\approx0.7$ LRG survey with the lower redshift LRG samples,
the overall cross--correlation result is now as consistent with a null detection 
as it is with the standard $\Lambda$CDM model for both \textit{W}-band  and ILC data.
For the ILC map, the significance of rejecting the standard model is 
$\approx2\sigma$ whereas the result is only $0.5\sigma$ away from the zero 
correlation hypothesis.

\item Given the results of the rotation test on the SDSS and 2SLAQ LRG samples,
the support these give to the standard ISW  model in the combined sample may
have even less statistical weight than indicated above.

\item There is a possibility that the absence of the ISW correlation in the high
redshift dataset is due to evolution of the dark energy equation of state.
However, we regard it as unlikely that evolution could take place over the short
redshift interval between the 2SLAQ and AA$\Omega$ datasets. It is more
plausible that the differences between the redshift bins are purely statistical,
particularly given the rotation test results. We note that the individual
positive detections that we have discussed are only marginally statistically
significant and the combined ILC dataset is more consistent with zero than with
the standard model prediction.

\item If the ISW effect was generally absent then the impact on cosmology would
be large because this would be strong evidence against an accelerating Universe.
This would therefore argue against a significant role for a cosmological constant or
dark energy in the Universe. Moreover, the absence of ISW would also argue
against any modified gravity model which produced acceleration. The model which
would be heavily favoured would be an Einstein-de Sitter model with
$\Omega_{\rm{m}}=1$. However, if such a model had a critical density of exotic,
CDM  particles then there might be a contradiction with the high baryon
densities in rich galaxy clusters such as Coma. This rich cluster `baryon
catastrophe' has previously argued against a high CDM density because starting
from $\Omega_b/\Omega_{\rm{m}}\approx0.03$, it was difficult to understand in a
hierarchical model how to produce a $5\times$ bigger baryon fraction in rich
galaxy clusters \citep{White93}.

\item It is therefore important to repeat the LRG measurements made here, now in the
Southern sky. One opportunity to do this will arise from the new ESO imaging
surveys in the South which are about to start, the VST ATLAS and the VISTA
Hemisphere Survey. If the results we have found here are repeated then there
could be significant consequences for cosmology.
 \end{itemize}

\section*{Acknowledgements}
US acknowledges financial support from the Institute for the 
Promotion of Teaching Science and Technology (IPST) of The Royal 
Thai Government. We thank Douglas Scott for useful discussion and
comments. We thank all the present and former staff of the 
Anglo--Australian Observatory for their work in building and 
operating the 2dF and AAOmega facility.

Funding for the SDSS and SDSS-II has been provided by the Alfred
P. Sloan Foundation, the Participating Institutions, the National
Science Foundation, the U.S. Department of Energy, the National
Aeronautics and Space Administration, the Japanese Monbukagakusho, the
Max Planck Society, and the Higher Education Funding Council for
England. The SDSS Web Site is {\tt http://www.sdss.org/}.

The SDSS is managed by the Astrophysical Research Consortium for the
Participating Institutions. The Participating Institutions are the
American Museum of Natural History, Astrophysical Institute Potsdam,
University of Basel, Cambridge University, Case Western Reserve
University, University of Chicago, Drexel University, Fermilab, the
Institute for Advanced Study, the Japan Participation Group, Johns
Hopkins University, the Joint Institute for Nuclear Astrophysics, the
Kavli Institute for Particle Astrophysics and Cosmology, the Korean
Scientist Group, the Chinese Academy of Sciences (LAMOST), Los Alamos
National Laboratory, the Max-Planck-Institute for Astronomy (MPIA),
the Max-Planck-Institute for Astrophysics (MPA), New Mexico State
University, Ohio State University, University of Pittsburgh,
University of Portsmouth, Princeton University, the United States
Naval Observatory, and the University of Washington.

\bibliographystyle{mn2e}
\bibliography{isw}

\label{lastpage}
\end{document}

%% file: format.tex
\def \zbar {\bar{\rm z}}

\def \Omm {\Omega_{\rm m}}

\def \hmpc{~\;h^{-1}~{\rm Mpc}}

\def \gsim { \lower .75ex \hbox{$\sim$} \llap{\raise .27ex \hbox{$>$}} }
\def \lsim { \lower .75ex \hbox{$\sim$} \llap{\raise .27ex \hbox{$<$}} }
\def \deg {^{\circ}}

\def \WMAP5 {{\it WMAP5}}

\def \W {{\textit{W }}}
\def \V {{\it V }}
\def \Q {{\it Q }}